\documentclass[journal]{IEEEtran}

\usepackage{cite}
\usepackage{url}

\usepackage{textcomp}
\usepackage{xcolor}
\usepackage{hyperref}

\usepackage{booktabs}	
\usepackage{diagbox}    
\usepackage{multirow}   

\usepackage{verbatim}	

\usepackage{amsmath,amssymb,amsfonts,graphicx}
\usepackage{epstopdf}

\newtheorem{definition}{Definition}[section]
\newtheorem{property}{Property}[section]

\newtheorem{lemma}{Lemma}  

\usepackage{graphicx,epstopdf,algpseudocode,caption,url}   
\usepackage[ruled,linesnumbered]{algorithm2e}

\usepackage{amssymb}
\usepackage{hyperref}

\usepackage[switch]{lineno}   

\ifCLASSINFOpdf

\else

\fi

\begin{document}

\title{TargetUM: Targeted High-Utility Itemset Querying}

\author{Jinbao Miao, Shicheng Wan, Wensheng Gan,~\IEEEmembership{Member,~IEEE,} Jiayi Sun, and Jiahui Chen,~\IEEEmembership{Member,~IEEE,}  
	

	\thanks{This work was partially supported by the National Natural Science Foundation of China (Grant Nos. 62002136 and 61902079), Guangzhou Basic and Applied Basic Research Foundation (Grant Nos. 202102020277 and 202102020928). (Corresponding author: Wensheng Gan)}

	\thanks{Jinbao Miao, Wensheng Gan, and Jiayi Sun are with the College of Cyber Security, Jinan University, Guangzhou 510632, China. (E-mail: osjbmiao@gmail.com, wsgan001@gmail.com, jiayisun01@gmail.com)}
	
	\thanks{Shicheng Wan and Jiahui Chen are with the School of Computers, Guangdong University of Technology, Guangzhou 510006, China. (E-mail: scwan1998@gmail.com, csjhchen@gmail.com)} 
}

\maketitle


\begin{abstract}

Traditional high-utility itemset mining (HUIM) aims to determine all high-utility itemsets (HUIs) that satisfy the minimum utility threshold (\textit{minUtil}) in transaction databases. However, in most applications, not all HUIs are interesting because only specific parts are required. Thus, targeted mining based on user preferences is more important than traditional mining tasks. This paper is the first to propose a target-based HUIM problem and to provide a clear formulation of the targeted utility mining task in a quantitative transaction database. A tree-based algorithm known as \textbf{T}arget-based high-\textbf{U}tility ite\textbf{M}set querying using (\textbf{TargetUM}) is proposed. The algorithm uses a lexicographic querying tree and three effective pruning strategies to improve the mining efficiency. We implemented experimental validation on several real and synthetic databases, and the results demonstrate that the performance of \textbf{TargetUM} is satisfactory, complete, and correct. Finally, owing to the lexicographic querying tree, the database no longer needs to be scanned repeatedly for multiple queries.
\end{abstract}


\IEEEpeerreviewmaketitle

\section{Introduction}
\label{sec:introduction}

With the rapid development of information technology and database management systems \cite{mccarthy1989architecture, stonebraker1991postgres}, extremely large amounts of raw data are produced daily. Massive useful messages are hidden in these big data. The discovery of valuable information and its utilization has emerged as an important topic. For example, customers usually buy keyboards and mouses after purchasing computers. Following this, other potential requirements exist, such as a pair of headphones and a suitable chair. However, customers may not gather these items into baskets simultaneously. Thus, we cannot intuitively observe the relationship among these goods. If the purchase records are analyzed according to data mining technologies \cite{aggarwal2014frequent}, the results will aid in understanding the knowledge that is hidden in rich data. For example, clerks can place headphones or chairs between computers and keyboards to improve the sale volume. In the past several decades, researchers have proposed hundreds of data mining algorithms that rely on different data formats in many applications and various domains. Among these, one traditional data mining technology is known as frequent pattern mining (FPM) \cite{agrawal1994fast, han2000mining}, and the Apriori \cite{agrawal1994fast} algorithm is the most famous and earliest algorithm. Apriori discovers interesting patterns based on frequency and confidence metrics. Since then, many other algorithms \cite{fournier2017survey, gan2017data} have been proposed for mining interesting patterns.

However, FPM algorithms exhibit a fatal flaw in that their results may be frequent but result in low profit. The frequency of a pattern may not be a sufficient indicator of interest as it only reflects the number of transactions that contain the pattern in a database. In fact, other important factors (e.g., weight, unit profit, and risk) often need to be considered in real-life applications. As a simple example, diamonds definitely have a lower sale volume than pencils. If only the frequency metric is considered, diamonds exhibit an unpromising pattern owing to their low support. Thus, they are apparently not a wise suggestion for retailers. To address this problem, inspired by economic knowledge, scholars have proposed a novel concept known as utility, and data mining technologies based on utility metrics have been presented in the form of utility-oriented pattern mining (UPM) \cite{gan2021survey}. Each pattern owns a unit utility (e.g., unit profit) and appears once or more in each transaction (e.g., purchase quantity). The UPM field has been developing for more than 20 years, and many algorithms have been designed to make the mining process more efficient, such as list-based algorithms (e.g., HUI-Miner \cite{liu2012mining}, FHM \cite{fournier2014fhm}, and FHN \cite{lin2016fhn}), tree-based algorithms (e.g., IHUP \cite{ahmed2009efficient}, UP-Growth \cite{tseng2012efficient}, and MU-Growth \cite{yun2014high}), projection-based algorithms (e.g., EFIM \cite{zida2017efim}), and other algorithms. In UPM, a pattern is referred to as a high-utility pattern (HUP) if its utility value is no less than a user-specified minimum utility threshold (abbreviated as \textit{minUtil}, denoted as $\sigma$). In general, UPM algorithms are more complex than FPM. For example, smart-phones and headphones can both generate high profits. However, most people buy these items separately instead of together because of their high prices. This indicates that selling a combination of goods may result in less profit than when selling them separately. However, if the goods are a computer and keyboard, the result will be the opposite. Therefore, utility is neither monotonic nor anti-monotonic, such as frequency. To address this limitation, Liu \textit{et al.} \cite{liu2005two} proposed an overestimated concept known as transaction-weighted utilization (abbreviated as \textit{TWU}), which has a downward closure property (i.e., anti-monotonic) \cite{agrawal1994fast}. This means that if a subset of a pattern is less than the threshold, all supersets thereof will not meet the condition either. Subsequently, all UPM algorithms adopt the \textit{TWU} concept, optimized data storage structures, and effective pruning strategies.

Nevertheless, most of the above algorithms cannot deal with target-oriented mining tasks. The users input a \textit{minUtil} threshold, following which the algorithms return all patterns with utility values that are higher than the threshold. That is, traditional UPM technologies aim to offer wide suggestions; they do not allow users to perform targeted queries. For example, shareholders simply wish to know when their own stocks will rise to the expected price, so that they can sell stocks in time, merchants often consider how to sell their stockpiled goods as soon as possible, and customers usually have a certain shopping list before going to the supermarket, which helps them to save time when purchasing goods. Thus, users are usually not really interested in obtaining the overall  information. Researchers have defined this interesting task as target-oriented pattern mining \cite{kubat2003itemset}. To date, several studies have been conducted on target-based pattern mining, such as target-oriented frequent itemset querying \cite{shabtay2018guided}, target-based association rule mining \cite{abeysinghe2017query, fournier2013meit}, and targeted sequential pattern querying \cite{chand2012target, chueh2010mining, zhang2021tusq}. As stated previously, customers are usually not interested in receiving all discount messages according to the recommender system. More precisely, they may prefer to gain knowledge on specific goods they need and whether these are on sale. Until recently, there has been a lack of knowledge and efficient technologies for target-based pattern mining. In this paper, we first formulate a new problem, known as target-based high-utility itemset mining (THUIM), to apply querying technology more extensively. In contrast to existing target-based algorithms, the novel task of targeted utility-oriented itemset querying focuses on extracting not only the results containing the predefined target, but also those with high utility values with respect to a specified $\sigma$.

Several specific challenges of THUIM need to be addressed. First, the utility does not naturally follow the anti-monotonic property, indicating that frequency-based approaches cannot be adopted directly. Second, traditional FPM and UPM algorithms generate numerous unpromising candidates. Therefore, it is difficult to discover targeted queries without tight upper bounds of utility, particularly in large-scale databases. Third, effectively saving on memory is an inescapable challenge. Thus, the design of a compact data structure is also required. Finally, there is an urgent requirement for algorithms that can process massive databases efficiently and provide acceptable querying performance. 

To this end, we propose the novel \textbf{Target}-based high-\textbf{U} utility ite\textbf{M}set querying algorithm (abbreviated as \textbf{TargetUM}). The main contributions of our study are as follows:

\begin{itemize}
	\item  To the best of our knowledge, this is the first algorithm that incorporates the concept of utility into target-based utility mining. We introduce several key definitions and formulate the problem of discovering the desired set of utility-driven target queries.
	
	\item  We adopt the utility-list structure to generate HUIs directly, which can prevent the production of redundant candidates. Furthermore, we propose a utility-based trie tree to query target itemsets conveniently and to improve the mining efficiency.
	
	\item  For further improvement in the efficiency, we use several upper bounds and pruning strategies to accelerate the calculation process.
	
	\item  We conduct experiments on several datasets to demonstrate the effectiveness and efficiency of TargetUM for mining all desired queries with a user specified \textit{minUtil}.
	
\end{itemize}

The remainder of this paper is organized as follows. Related work is presented in Section \ref{sec:relatedwork}. Thereafter, key preliminaries are described in Section \ref{sec:preliminaries}. Section \ref{sec:algorithm} presents the details of the designed TargetUM algorithm. The experiments and a comprehensive analysis are outlined in Section \ref{sec:experiments}. Finally, the conclusions and future work are presented in Section \ref{sec:conclusion}.

\section{Related Work}
\label{sec:relatedwork}

In this section, we briefly review several studies on traditional frequent itemset mining (FIM), HUIM, and target pattern querying.

\subsection{Frequent itemset mining}

Given a transaction database, FIM aims to identify the itemsets that appear frequently. FIM and its extended technologies have been applied extensively in numerous real-life domains \cite{aggarwal2014frequent, agrawal1994fast, han2000mining, zaki2000scalable}. Retailers use frequent itemsets to promote frequently purchased goods in market basket analysis. The most representative work is Apriori \cite{agrawal1994fast}, which is a level-wise algorithm. Although its drawback is obvious, namely that it generates too many candidates and has to scan the database repeatedly, it has been possible to derive many Apriori-like algorithms \cite{liu2005two, pasquier1999discovering, szathmary2007towards}. Subsequently, FP-Growth \cite{han2000mining} established a new tree-based algorithm. According to the FP-tree data structure, the database only needs to be scanned twice and a highly compact data tree is constructed. The FP-tree stores all key information regarding frequent super-itemsets and uses a head-table to record the frequent items. Therefore, FP-Growth can conveniently obtain all interesting itemsets and it requires fewer resources than Apriori. In general, numerous investigators have improved FIM algorithms in recent decades \cite{aryabarzan2021neclatclosed, aryabarzan2018negfin, pei2007h, zaki2000scalable}. However, as explained in Section \ref{sec:introduction}, most FIM algorithms do not consider information regarding the unit interesting factor of items. Thus, many frequent itemsets with low interest will be identified; for example, many rare itemsets with high profits may be discarded. To address this issue, UPM \cite{gan2021survey} has attracted considerable attention. In a sense, utility refers to the profit/risk that an item can bring.

\subsection{High-utility itemset mining}

HUIM is a subfield of UPM. Since the first HUIM algorithm Two-Phase \cite{liu2005two} was proposed, HUIM technologies have been used extensively in many practical applications, such as user behavior analysis \cite{shie2013mining}, website click-stream analysis \cite{chu2008efficient}, and cross-marketing analysis \cite{yen2007mining}. Two distinct types of algorithms exist in the HUIM field (two-phase and single-phase models). In the well-known two-phase model, the mining process (referred to as Phase I) selects HUIs after computing all of the candidates (referred to as Phase II). Two-Phase incorporates an upper bound known as transaction-weighted utilization (\textit{TWU}), which has a downward closure property, meaning that super-itemsets of an itemset cannot be HUIs if their \textit{TWU} values are less than \textit{minUtil}. Other HUIM algorithms include \textit{TWU} to prune unpromising items/itemsets easily, whereas these two-phase models compute HUIs that need to scan the database at least twice. Hence, IHUP \cite{ahmed2009efficient} incorporates a compact searching tree similar to an FP-tree \cite{han2000mining}. The new data structure saves more runtime and memory compared to Two-Phase. However, IHUP uses an inaccurate upper bound so that many unpromising super-itemsets are retained. UP-Growth \cite{tseng2010up} and UP-Growth+ \cite{tseng2012efficient} successfully solved this issue by designing a complete search tree structure to discover all real HUIs.

In general, a tree data structure can avoid the database being scanned multiple times. However, when dealing with huge amounts of data, the tree will be more complex and will run out of memory. Liu \textit{et al.} \cite{liu2012mining} proposed a breakthrough algorithm known as HUI-Miner, which is classified as a single-phase model. Compared to two-phase algorithms, the single-phase model calculates HUIs without generating candidates. HUI-Miner uses a novel data structure known as a utility-list, which can avoid the problem of numerous candidates being generated. Moreover, a new upper bound depending on the remaining utility can filter out uninteresting itemsets in advance. Subsequently, other list-based algorithms that perform better than HUI-Miner have been proposed. For example, FHM \cite{fournier2014fhm} uses a new list structure known as the estimated utility co-occurrence structure to reduce the expense of intersection/join operations; FHN \cite{lin2016fhn} is an improved version of FHM that can mine HUIs with negative utility values; the ULB-Miner algorithm \cite{duong2018efficient} includes an improved data structure known as the utility-list buffer to reduce the memory consumption and accelerate the join process; and the state-of-the-art TopHUI \cite{gan2020tophui} addresses the top-$k$ utility mining problem without setting \textit{minUtil}. Other new data structures and mining algorithms have also been proposed, such as TKO \cite{tseng2015efficient}, KHMC \cite{duong2016efficient}, EFIM \cite{zida2017efim}, and HUOPM \cite{gan2019huopm}. Further details regarding UPM can be found in Ref. \cite{gan2018privacy,gan2021survey}.

\subsection{Target pattern querying}

Most pattern mining algorithms are designed to discover the complete itemsets in an entire database. In reality, not all of these patterns are interesting. However, users focus only on a part of the mining results. Therefore, target pattern querying \cite{kubat2003itemset}, which is similar to an interactive query method, was introduced. In this method, users offer the target set and the system searches for related objects based on the user input. Users can repeatedly query their target patterns from the mining results. Several target-oriented querying approaches have been proposed. In 2003, Kubat \textit{et al.} \cite{kubat2003itemset} introduced the target-oriented querying task and designed an approach to handle the specialized queries of users in a transaction database. They designed Itemset-Tree (IT), which differs from the FP-tree structure. The insertion of new transactions to update nodes incrementally provides an efficient querying function. However, the IT data structure consumes a large amount of memory owing to its poor scalability. Fournier-Viger \textit{et al.} \cite{fournier2013meit} designed an improved data structure known as the memory-efficient itemset tree (MEIT) to address this issue. The new structure uses a node compression mechanism to reduce the size of the IT node efficiently. The experiments demonstrated that the new data structure is up to 43\% smaller than IT on average in terms of memory consumption. As an FP-tree-based algorithm, guided FP-growth (GFP-growth) \cite{shabtay2018guided} was proposed. It can determine the support of a given large list of itemsets (that are regarded as targets) using the target itemset tree. Several target-based approaches have also been developed to handle sequence data based on target query constraints. Chueh \textit{et al.} \cite{chueh2010mining} presented an algorithm that uses time intervals to reverse the original sequences and to determine target sequential patterns. Chand \textit{et al.} \cite{chand2012target} extended targeted sequential pattern mining with monetary constraints. However, the aforementioned sequential mining algorithms focus on frequency metrics. The TUSQ algorithm \cite{zhang2021tusq} is the first to incorporate the utility concept into target pattern querying. Its motivation is that utility-driven sequential pattern mining algorithms often obtain several useless patterns owing to exhaustive results. At present, mining algorithms with target-querying constraints have been applied in various applications \cite{abeysinghe2017query, abeysinghe2018query}.

To date, no research has integrated utility with target itemset querying in quantitative databases. The existing HUIM algorithms return a large number of HUIs, which may waste time for users in discovering the targeted special results. This motivated us to design an efficient target-querying approach for mining HUIs from quantitative transaction databases.

\section{Preliminaries and Problem Statement}
\label{sec:preliminaries}

In this section, we first briefly introduce the basic concepts relating to HUIM. Most notations and definitions are provided in \cite{liu2012mining,lin2016fhn}, and we formulate a new problem statement and the parameter requirements thereof.

\subsection{Basic preliminaries}

As defined in previous works, $I$ = $\{i_1$, $i_2$, $i_3$, $\ldots$, $i_n\}$ is a set of $n$ distinct items in a transaction database ($\mathcal{D}$). $\mathcal{D}$ is mainly composed of several transactions, which are defined as $\mathcal{D}$ = $\{T_1$, $T_2$, $T_3$, $\ldots$, $T_m\}$. We use $T_{tid}$ to mark each transaction with a unique ID. Each transaction consists of many different items. In particular, an itemset is known as an $l$-itemset (1 $\le$ $l$ $\le$ $n$) if it includes $l$ items, and any itemset is a subset of $I$. Each item $i_{\ell}$, where $1 \leq \ell \leq m$ has a non-negative quantity $iu(i_{\ell}$, $t_j)$, which represents the occurring quantity (known as \textit{internal utility}). The unit utility (referred to as the \textit{external utility}) of $i_{\ell}$ is defined as $eu(i_{\ell})$.

\begin{table}[!h]
	\centering
	\caption{A simple example}
	\label{Tab:db}
	\centerline{(a) A quantitative transaction database}
	
	\begin{tabular}{ccc}
		\hline
		\textbf{TID} & \textbf{Transaction} & \textbf{Quantity}  \\ \hline
		$T_1$    	 & $\{b,c,e,g,h\}$  	& $\{1,2,2,1,1\}$ \\
		$T_2$    	 & $\{a,c,f,g\}$  		& $\{2,1,2,4\}$	  \\
		$T_3$    	 & $\{b,c,e,h\}$    	& $\{5,2,3,4\}$   \\
		$T_4$    	 & $\{a,c,d,e,g\}$ 	    & $\{2,1,3,1,2\}$ \\
		$T_5$    	 & $\{c,e\}$    		& $\{3,4\}$		  \\
		$T_6$    	 & $\{g,h\}$     		& $\{1,1\}$	      \\
		\hline
	\end{tabular}
	
	\vspace{3mm}
	
	\centerline{(b) Utility of single items}
	\label{Tab:profit}
	\begin{tabular}{lllllllll}
		\hline
		\textbf{Item} 	 & $a$   & $b$ 	 & $c$   & $d$   & $e$   & $f$   & $g$   & $h$   \\ \hline
		\textbf{Utility} & \$3 & \$2 & \$1 & \$5 & \$4 & \$2 & \$2 & \$1 \\ \hline
	\end{tabular}	
\end{table}

\begin{definition}
	\rm  (\textit{Utility of item}) In each transaction $T_j$, the utility of item $i_\imath$ is denoted as $u(i_\imath, T_j)$, where $u(i_\imath, T_j)$ = $eu(i_\imath)$ $\times$ $iu(i_\imath, T_j)$. According to the previous formula, the utility of item $i_\imath$ in database $u(i_\imath)$ consists of $\sum_{i_\imath \in T_j \land T_j \subseteq \mathcal{D}}$$u(i_\imath, T_j)$ = $\sum_{T_j \in \mathcal{D}}$($eu(i_\imath)$ $\times$ $iu(i_\imath, T_j)$).
\end{definition}

For example, in the second transaction in Table \ref{Tab:db}, $u(c, T_2)$ = $eu(c)$ $\times$ $iu(c, T_2)$ = \$1 $\times$ 1 = \$1. When considering the third transaction, $u(c, T_3)$ = $eu(c)$ $\times$ $iu(c, T_3)$ = \$1 $\times$ 2 = \$2. Therefore, the total utility of item $c$ in $ \mathcal{D}$ is $u(c)$ = $T_1$ + $T_2$ + $T_3$ + $T_4$ + $T_5$ = \$9.

\begin{definition}
	\rm  (\textit{Utility of an itemset}) The utility of an itemset $X_i$ in a transaction $T_j$ is denoted as $u(X_i$, $T_j)$ = $\sum_{i_\imath \in X_i \wedge X_i \subseteq T_j}u(i_\imath, T_j)$. Similar to $u(i_\imath)$, the utility of an itemset in the database is defined as $u(X_i)$ = $\sum_{X_i \subseteq T_j \land T_j \subseteq \mathcal{D}}u(X_i, T_j)$ = $\sum_{X_i \subseteq T_j \wedge T_j \in \mathcal{D}}u(X_i,  T_j)$ = $\sum_{X_i \subseteq T_j \wedge T_j\in \mathcal{D}}\sum_{i_\imath \in X_i}u(i_\imath,  T_j)$.
\end{definition}

For example, in Table \ref{Tab:db}, $u(\{c, e\},T_1)$ = $u(c,T_1)$ + $u(e,T_1)$ = \$1 $\times$ 2 + \$4 $\times$ 2 = \$10, $u(\{c, e\},T_3)$ = $u(c,T_3)$ + $u(e,T_3)$ = \$1 $\times$ 2 + \$4 $\times$ 3 = \$14, and $u(\{c, e\},T_4)$ = $u(c,T_4)$ + $u(e,T_4)$ = \$1 $\times$ 1 + \$4 $\times$ 1 = \$5. Furthermore, $u(\{c, e\})$ = $u(\{c, e\},T_1)$ + $u(\{c, e\},T_3)$ + $u(\{c, e\},T_4)$ + $u(\{c, e\},T_5)$ = \$10 + \$14 + \$5 + \$19 = \$48, and $u(\{g, h\})$ = $u(\{g, h\},T_1)$ + $u(\{g, h\},T_6)$ = \$3 + \$3 = \$6.

\begin{definition}
	\rm  (\textit{Utility of transaction}) The summed utility of all items in transaction $T_j$ is denoted as the utility of $T_j$, which is computed as $tu(T_j)$ = $\sum_{i_\imath \in T_j}u(i_\imath, T_j)$ where 1 $\le$ $j$ $\le$ $m$.
\end{definition}

For example, $tu(T_1)$ = $u(b, T_1)$ + $u(c, T_1)$ + $u(e, T_1)$ + $u(g, T_1)$ + $u(h, T_1)$ = \$2 $\times$ 1 + \$1 $\times$ 2 + \$4 $\times$ 2 + \$2 $\times$ 1 + \$1 $\times$ 1 = \$15, $tu(T_3)$ = $u(b, T_3)$ + $u(c, T_3)$ + $u(e, T_3)$ + $u(h, T_3)$ = \$2 $\times$ 5 + \$1 $\times$ 2 + \$ $\times$ 3 + \$1 $\times$ 4 = \$28, and the remaining utilities of transactions are presented in Table \ref{Tab:transaction_utility}.

\begin{table}[!h]
	\begin{center}
		\caption{Transaction utility}
		\label{Tab:transaction_utility}
		\begin{tabular}{lcccccc}
			\hline
			\textbf{TID} 		  & $T_1$ & $T_2$ & $T_3$ & $T_4$ & $T_5$ & $T_6$ \\ \hline
			\textbf{\textit{TU}}  & \$15  & \$19  & \$28  & \$30  & \$19  & \$3   \\ \hline
		\end{tabular}
	\end{center}
\end{table}

\begin{definition}
	\rm  (\textit{Transaction-weighted utilization}) Given an itemset $X_i$ belonging to $\mathcal{D}$, the TWU of $X_i$ is denoted as \textit{TWU}$(X_i)$, which determines whether $X_i$ is a potential HUI. This is the sum of the utilities of all transactions containing $X_i$ in $\mathcal{D}$, where \textit{TWU}$(X_i)$ = $\sum_{T_j \in \mathcal{D} \wedge X_i \subseteq T_j}$$tu(T_j)$. It is an upper bound on real utility. The details of the proof can be found in Ref. \cite{liu2012mining}.
\end{definition}

For example, in Table \ref{Tab:db}, \textit{TWU}$(a)$ = $(tu(T_2)$ + $tu(T_4))$ = \$19 + \$30 =\$49, \textit{TWU}$(c)$ = $tu(T_1)$ + $tu(T_2)$ + $tu(T_3)$ + $tu(T_4)$ + $tu(T_5)$ = \$15 + \$19 + \$28 + \$30 + \$19 = \$111, and the \textit{TWU} values of all items are listed in Table \ref{Tab:twu}.

\begin{table}[!h]
	\begin{center}
		\caption{Transaction-weighted utility}
		\label{Tab:twu}
		
		\centerline{(a)}	
		\begin{tabular}{lcccccccc}
			\hline
			\textbf{item} 		  & $a$  & $b$  & $c$   & $d$  & $e$  & $f$  & $g$  & $h$  \\ \hline
			\textbf{\textit{TWU}} & \$49 & \$43 & \$111 & \$30 & \$92 & \$19 & \$67 & \$46 \\ \hline
		\end{tabular}
		
		\vspace{3mm}
		\centerline{(b)}  	
		\begin{tabular}{lcccccccc}
			\hline
			\textbf{item} 		  & $f$   & $d$   & $b$  & $h$  & $a$  & $g$  & $e$ & $c$  \\ \hline
			\textbf{\textit{TWU}} & \$19 & \$30 & \$43 & \$46 & \$49 & \$67 & \$92 & \$111 \\ \hline
		\end{tabular}
	\end{center}
\end{table}

\begin{definition}
	\rm (\textit{Utility-list}) The utility-list data structure \cite{liu2012mining} of itemset $X_i$ consists of several tuples (including \textit{tid}, \textit{iutil}, and \textit{rutil} for each transaction $T_{tid}$ that includes $X_i$). The term \textit{iutil} is the total utility of $X_i$ in $T_{tid}$, and the \textit{rutil} element is denoted by $\sum_{x_j \in T_{tid} \land x_j \not\in X_i}$$u(x_j, T_{tid})$ \cite{liu2012mining,lin2016fhn}. 
\end{definition}

For example, the utility-list of $\{g\}$ is $\{T_1, \$2, \$1\}$, $\{T_2, \$8, 0\}$, $\{T_4, \$4, 0\}$ and $\{T_6, \$2, \$1\}$. The utility-list of $\{e\}$ is $\{T_1, \$8, \$3\}$, $\{T_3, \$12, \$4\}$, $\{T_4, \$4, \$4\}$ and $\{T_5, \$16, 0\}$. Therefore, the utility-list of $\{e, g\}$ is $\{T_1, \$10, \$1\}$ and $\{T_4, \$8, 0\}$.

\begin{definition}
	\rm  (\textit{HUI}) The minimum utility threshold (abbreviated as \textit{minUtil}) is a special constraint that is predefined by users. An itemset $X_i$ is known as an HUI if its utility value is no less than \textit{minUtil} $\sigma$ (i.e., $u(X_i) \ge \sigma$). Otherwise, we suppose that $X_i$ is an uninteresting itemset and ignore it. In our work, we require two thresholds to select HUIs from target itemsets: \textit{minUtil} is denoted as $\sigma$, whereas target \textit{minUtil} is defined as $\xi$.
\end{definition}

For example, if we suppose that $\sigma$ is \$30, we can observe all high-utility itemsets from Table \ref{Tab:high_utility_itemsets}. In all 1-itemsets, only $e$ is an HUI because $u(e)$ = \$8 + \$12 + \$4 + \$16 = \$40 according to Table \ref{Tab:db}.

\begin{table}[!h]
	\begin{center}
		\caption{High-utility itemsets w.r.t \textit{minUtil} = \$30}
		\label{Tab:high_utility_itemsets}
		\begin{tabular}{cc|cc}
			\hline
			\textbf{itemset}  & \textbf{utility} & \textbf{itemset} & \textbf{utility} \\ \hline
			$\{d, a, g, e, c\}$ & \$30 & $\{b, e\}$ 	  & \$32 \\ \hline
			$\{b, e, c\}$ 	    & \$36 & $\{b, h, e\}$ 	  & \$37 \\ \hline
			$\{e\}$ 			& \$40 & $\{b, h, e, c\}$ & \$41 \\ \hline
			$\{e, c\}$ 		    & \$48 & - 			      & - 	 \\ \hline
		\end{tabular}
	\end{center}
\end{table}

\begin{definition}
	\rm  (\textit{Target high-utility itemset}) In practice, target itemsets are a user-specified set $T^\prime$. The term target means that these itemsets are of special interest. It should be noted that $\mid$$T^\prime$$\mid$ is no less than 1. The formulaic definition is $T^\prime$ = $<$$i_1$, $i_2$, $\dots$, $i_\imath$, $\dots$, $i_p$$>$ ($\imath \in [1, p] \wedge i_\imath \in I)$. Suppose that an itemset $X_i$ is an HUI, and $T^\prime$ is a given target itemset; if $\forall$$\imath$ that $i_\imath \in \mathcal{D}$ and $i_\imath \in T^\prime$, where $i_\imath$ $\in$ $X_i$, we regard $X_i$ as a target high-utility itemset (abbreviated as \textit{THUI}), in which $u(X_i)$ $>$ $\xi$ and $T^ \prime \subseteq X_i$.
\end{definition}

\subsection{Problem statement}

At present, utility mining algorithms generally discover all high-utility patterns with utility values of no less than $\sigma$. However, in real applications, not all high-utility patterns are required by the user. For example, the user may wish to determine the high-utility patterns that include a certain item or itemset. We refer to such patterns as target patterns regarding a certain item or itemset. Thus far, we have introduced the related basic concepts and preliminaries. We formulate the addressed problem in detail below. 

\textbf{Problem statement}: The goal of THUIM is to identify all THUIs with two pre-specified minimum utility thresholds by the user. Specifically, the problem of THUI mining can be interpreted as discovering all itemsets that contain target items and have a utility value of no less than $\sigma$ and $\xi$.

For example, according to Table \ref{Tab:high_utility_itemsets}, if we set $T^\prime$ = $\{b, e\}$ and $\xi$ = \$30, the THUIs are $\{b, e, c\}$ (= \$36), $\{b, h, e\}$ (= \$37), and $\{b, h, e, c\}$ (= \$41), respectively. Moreover, if we set $T^\prime$ = $\{e, c\}$, we obtain $\{d, a, g, e, c\}$ (= \$30), $\{b, h, e, c\}$ (= \$41), $\{b, e, c\}$ (= \$36), and $\{e, c\}$ (= \$48), respectively. Obviously, $\xi$ can also be lower than $\sigma$, depending on the user's requirement.

\section{TargetUM Algorithm}
\label{sec:algorithm}

In this section, we present the target-based itemset querying algorithm known as TargetUM. The storage of THUIs mainly depends on the trie-tree, and eventually, the target itemsets are identified by on-the-fly querying in the tree. TargetUM can also flexibly query different target itemsets repeatedly with the same $\sigma$. In the following section, we discuss pruning strategies and outline the detailed steps of TargetUM.

\subsection{Target pattern tree}

\begin{figure}[!htbp]
	\centering
	\includegraphics[scale=0.34]{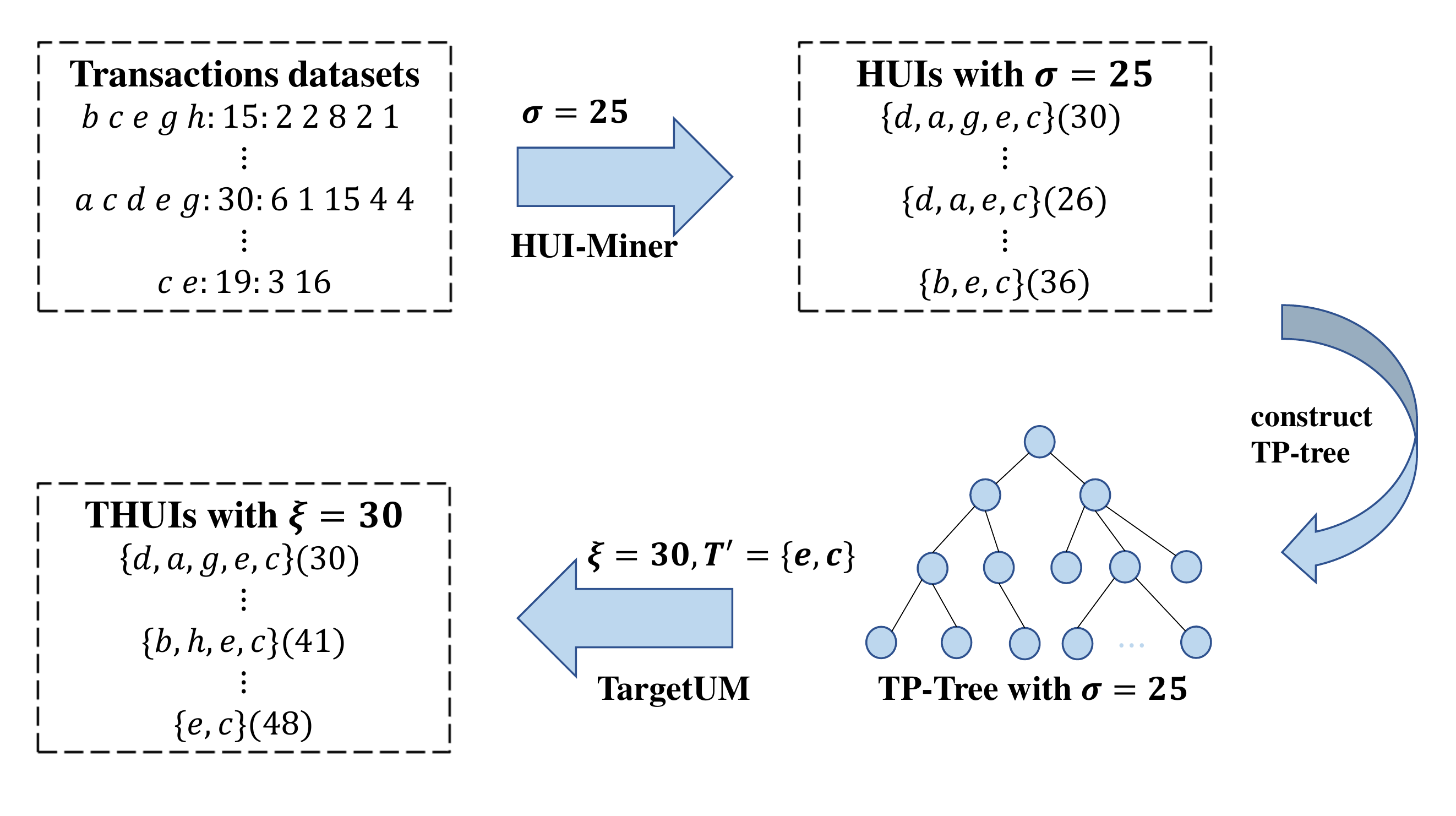}
	\caption{The TargetUM framework.}
	\label{fig:TargetUM_framework}
\end{figure}

As a key part of the TargetUM algorithm, the target pattern tree (TP-tree) structure is actually a trie tree. In an earlier study, MEIT \cite{fournier2013meit} used a compact node compression mechanism to reduce the memory usage. In contrast to MEIT, TP-tree stores the complete HUIs after completing the mining process on the entire database instead of each transaction, which takes full advantage of the highly compact character of the trie tree. Finally, all HUIs containing the target items or itemsets can be identified. The flowchart of the proposed TargetUM framework is illustrated in Fig. \ref{fig:TargetUM_framework}.  

\begin{figure*}[!htbp]
	\centering
	\includegraphics[scale=0.54]{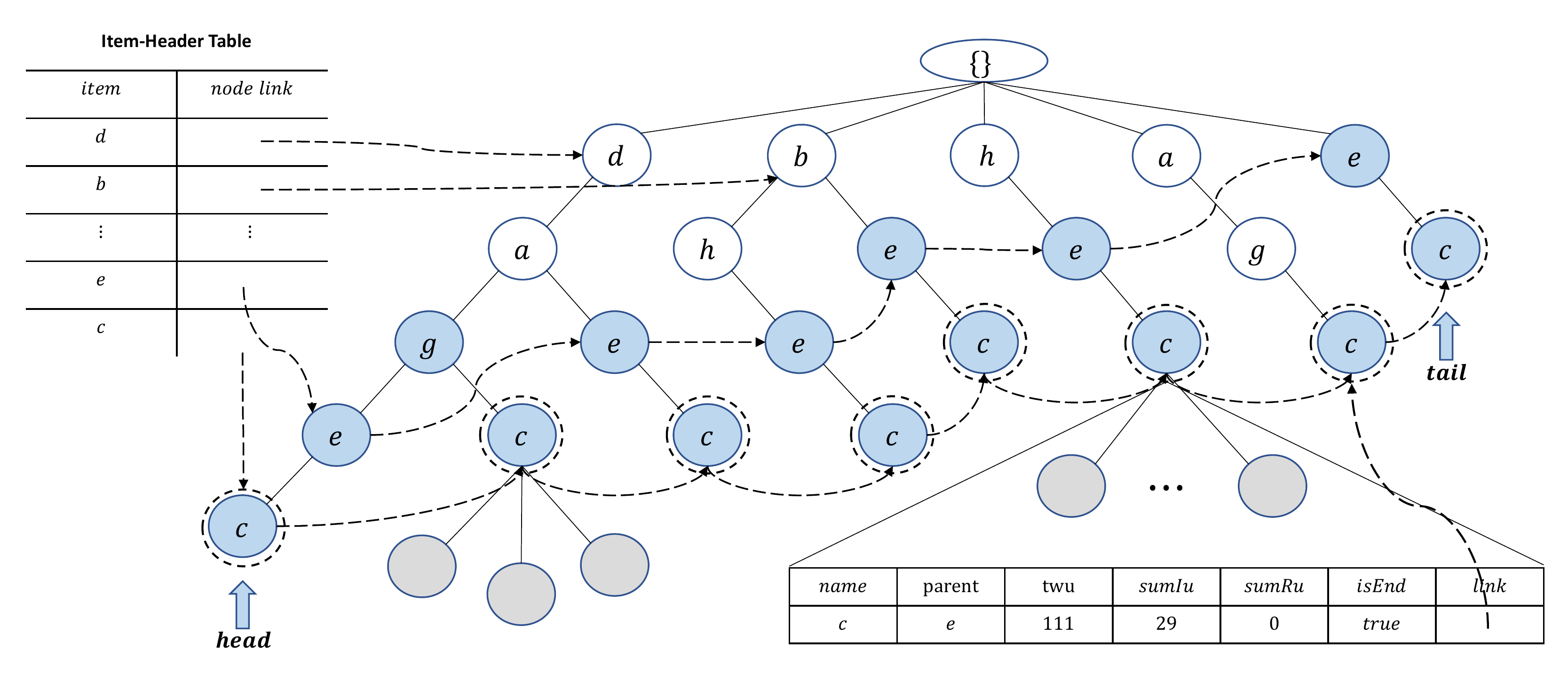}
	\caption{The constructed TP-tree with $\sigma$ = \$25.}
	\label{fig:TP_tree_node}
\end{figure*}

\begin{definition}
	\rm (\textit{Target pattern tree}) Each node in the TP-tree contains \textit{name}, \textit{parent}, \textit{twu}, \textit{sumIu}, \textit{sumRu}, \textit{isEnd}, and \textit{link}. As indicated in Fig. \ref{fig:TP_tree_node}, \textit{name} represents each item and \textit{link} is the unique chain of each item. The \textit{children} and \textit{parent} of a node in a TP-tree link other nodes and all of these are used to construct the trie. Obviously, all nodes in the branches are sorted by \textit{TWU}-ascending order. The \textit{twu}, \textit{sumIu}, and \textit{sumRu} are used to record the \textit{TWU}, total utility, and total remaining utility of each item/node, respectively. Finally, \textit{isEnd} indicates whether the current node is the last item of the HUI, because the TP-tree is always traversed from top to bottom.
\end{definition}

\begin{definition}
	\rm (\textit{Priority of items}) To facilitate the mining process in TargetUM, all of the items in each transaction are sorted by \textit{TWU}-ascending order, and each transaction is revised. Table \ref{Tab:twu}(b) presents the final status of all items.
\end{definition}

\begin{definition}
	\rm  (\textit{Remaining utility of itemset}) Given an itemset $X_i$ and a transaction $T_j$ with $X_i$ $\subseteq$ $T_j$, all items behind $X_i$ in $T_j$ are known as the remaining items ($T_j / X_i$). Subsequently, the remaining utility of $X_i$ in $T_j$ is denoted as $ru(X_i,T_j)$ = $\sum_{X_i\subseteq T_j \wedge x_\imath \in T_j/X_i}$$u(x_\imath, T_j)$ \cite{liu2012mining}.
\end{definition}

For example, in Table \ref{Tab:db}, if we set $X_i$ = $\{b, h\}$, $T_1$ / $X_i$ = $\{g, e, c\}$ and $T_3$ / $X_i$ = $\{e, c\}$. Therefore, $ru(X_i, T_1)$ = $u(g,T_1)$ + $u(e,T_1)$ + $u(c,T_1)$ = \$2 $\times$ 1 + \$4 $\times$ 2 + \$1 $\times$ 2 = \$12, and $ru(X_i, T_3)$ = $u(e,T_3)$ + $u(c,T_3)$ = \$4 $\times$ 3 + \$1 $\times$ 2 = \$14. When setting $\sigma$ = \$25, all HUPs can be identified using an existing HUIM algorithm, as indicated in Table \ref{Tab:five}.

\begin{table}[!h]
	\begin{center}
		\caption{High-utility patterns w.r.t $\sigma$ = \$25}
		\label{Tab:five}
		\begin{tabular}{c | c | c}
			\hline
			\bfseries{\textbf{HUI}} & \bfseries{\textbf{Utility}} & \bfseries{\textbf{Remaining utility}} \\ 
			\hline
			$\{d, a, g\}$ & $\{\$15, \$21, \$25\}$ & $\{\$15, \$9, \$5\}$ \\ \hline
			$\{b, e\}$ & $\{\$12, \$32\}$ & $\{\$31, \$4\}$ \\ \hline
			$\{d, a, g, e\}$ & $\{\$15, \$21, \$25, \$29\}$ & $\{\$15, \$9, \$5, \$1\}$ \\ \hline
			$\{b, e, c\}$ & $\{\$12, \$32, \$36\}$ & $\{\$31, \$4, 0\}$ \\ \hline
			$\{d, a, g, e, c\}$ & $\{\$15, \$21, \$25, \$29, \$30\}$ & $\{\$15, \$9, \$5, \$1, 0\}$ \\ \hline
			$\{h, e\}$ & $\{\$6, \$25\}$ & $\{\$28, \$4\}$ \\ \hline
			$\{d, a, g, c\}$ & $\{\$15, \$21, \$25, \$26\}$ & $\{\$15, \$9, \$5, 0\}$ \\ \hline
			$\{h, e, c\}$ & $\{\$6, \$25, \$29\}$ & $\{\$28, \$4, 0\}$ \\ \hline
			$\{d, a, e\}$ & $\{\$15, \$21, \$25\}$ & $\{\$15, \$9, \$1\}$ \\ \hline
			$\{a, g, c\}$ & $\{\$12, \$24, \$26\}$ & $\{\$18, \$6, 0\}$ \\ \hline
			$\{d, a, e, c\}$ & $\{\$15, \$21, \$25, \$26\}$ & $\{\$15, \$9, \$1, 0\}$ \\ \hline
			$\{e\}$ & $\{\$40\}$ & $\{\$8\}$ \\ \hline
			$\{b, h, e\}$ & $\{\$12, \$17, \$37\}$ & $\{\$31, \$26, \$4\}$ \\ \hline
			$\{e, c\}$ & $\{\$40, \$48\}$ & $\{\$8, 0\}$ \\ \hline
			$\{b, h, e, c\}$ & $\{\$12, \$17, \$37, \$41\}$ & $\{\$31, \$26, \$4, 0\}$ \\
			\hline
		\end{tabular}
	\end{center}
\end{table}

It should be noted that the mining process and TP-tree construction are executed simultaneously (i.e., every discovered HUI is immediately inserted into the TP-tree). Taking Table \ref{Tab:db} as an example, the details of how to construct a TP-tree are introduced in the following. 

Initially, the root node of the TP-tree is always empty. After determining the first HUI$_1$ $\{d, a, g\}$, the remaining utility values are also recorded. Subsequently, three nodes ($d$, $a$, and $g$) are inserted into the TP-tree. As the priority of nodes, $d$ is the first one. At this time, it will be verified whether there already exists a node $d$ that is directly connected to the root node. If so, HUI$_1$ shares a common prefix $<$$d$$>$ with the remaining items. Otherwise, a new node $d$ is created and the relative parameters are set as \textit{d.name} = $d$, \textit{d.parent} = \textit{root}, \textit{d.twu} = \$30, \textit{d.sumIu} = \$15, and \textit{d.sumRu} = \$15. As $d$ is not the last item in HUI$_1$, \textit{d.isEnd} is \textit{false}. An item header table of $d$ is created simultaneously. The function of the item header table can quickly locate the position of an item in the TP-tree. Because item $d$ is first added to the item header table, it is marked as a head node and tail node together. The remaining items/itemsets are processed and repeated using the same steps. Thus far, a branch $\{d, a, g, e, c\}$ of the TP-tree has been successfully created. Based on this branch, the method of inserting another HUI$_2$ $\{d, a, e\}$ is considered.

It can easily be observed that HUI$_1$ and HUI$_2$ share the common prefix $<$$d, $$>$; thus, HUI$_2$ creates a new branch from HUI$_1$. Owing to the characteristics of the TP-tree, the overall key information regarding $u(X)$ and $ru(X)$ of itemset $X$ is fixed in the same branch. Therefore, we only need to create a new node $e$, and set the remaining parameters as \textit{e.name} = $e$, \textit{e.parent} = $a$, \textit{e.twu} = \$92, \textit{g.sumIu} = \$25, \textit{g.sumRu} = \$1, and \textit{e.isEnd} = \textit{true}. Because item $e$ already exists in the item header table, we link the new node and then set the last $e$ as the new tail node directly. Similarly, the processes of the other HUIs follow the same steps. Finally, a complete TP-tree is constructed, as depicted in Fig. \ref{fig:TP_tree_node}.

\subsection{Search space for mining THUIs}

In this section, we explain the search space of a TP-tree constructed with HUIs. The TP-tree is an $n$-array tree rather than a binary tree. As indicated in Table \ref{Tab:twu}(b), all nodes/items are sorted by \textit{TWU}-ascending order. To reduce the searching consumption, TargetUM discovers target itemsets in a bottom-to-top manner. Several important properties and lemmas of the TP-tree are listed below.

\begin{property}
	\rm The \textit{TWU} of each node in the TP-tree is no less than its parent node.
\end{property}

\begin{property}
	\rm Each HUI is a unique path/branch of the TP-tree and this path starts with the root node.
\end{property}

\begin{property}
	\rm The insertion of an HUI into the TP-tree does not update the utility and remaining utility values of the existing nodes.
\end{property}

\begin{lemma} 
	\rm In the TP-tree, each node represents an item. Let $X^k$ be a $k$-itemset (node) and let its parent node be denoted by $X^{k-1}$. According to the TWU property \cite{liu2005two}, we obtain \textit{TWU}($X^k$) $\le$ \textit{TWU}($X^{k-1}$).
\end{lemma}

\textit{Proof}: $X^k$ is constructed by joining the pairs of $X^{k-1}$, and the transaction containing $X^k$ also will include $X^{k-1}$. Furthermore, the number of transactions containing $X^{k-1}$ is always higher than or equal to that of $X^k$. Therefore, considering the definition of \textit{TWU}, we obtain \textit{TWU}($X^k$) $\le$ \textit{TWU}($X^{k-1}$).

\begin{lemma} 
	\rm Every branch of the TP-tree is a unique path and all of the HUIs with $\sigma$ can be derived from the TP-tree. We can discover all THUIs with $\xi$ and $T^\prime$ from the TP-tree.
\end{lemma}

\textit{Proof}: In the TP-tree, each node except for the root only has one parent node. In the construction process of the TP-tree, each HUI is mapped to a unique path in the tree. Accordingly, the TP-tree contains all itemsets with $\sigma$ and provides the unique determination of each THUI.

\subsection{Efficient pruning strategies}

In the designed TP-tree, itemset $X^{k}$ is obtained by performing the join operation of the utility-lists of itemsets $X^{k-1}$. This not only effectively accelerates the candidate generation, but also saves time when querying targeted itemsets. In the preceding subsection, we described how to use the TP-tree to store key information and then to discover HUIs. In this subsection, we introduce several pruning strategies.

As mentioned in \cite{liu2005two, fournier2014fhm, zida2017efim}, \textit{TWU} is a tight upper bound that has three important properties (overestimation, anti-monotonicity, and pruning). Let $X_i$ and $X_j$ be two distinct itemsets. If $X_i \subset X_j$, \textit{TWU}($X_i$) $\ge$ \textit{TWU}($X_j$) because of the anti-monotonicity. The overestimation property means that the \textit{TWU} of an itemset is always higher than or equal to its real utility.

\textbf{Strategy 1}: Based on the anti-monotonicity and overestimation, it is clear that if \textit{TWU} of an itemset $X_i$ is less than \textit{minUtil} ($\sigma$ in this paper), $X_i$ will be a low-utility itemset as well as its supersets \cite{liu2005two}.


Let the sum of the utilities in the utility-list of an itemset $X_i$ be denoted as \textit{sumIu}($X_i$). If \textit{sumIu}($X_i$) is no less than \textit{minUtil}, $X_i$ is an HUI; otherwise, it is an uninteresting itemset. An item $x_j$ after $X_i$ in $T_j$ that can be appended to $X_i$ is known as the remaining item of $X_i$. Moreover, \textit{sumRu}($X_i$) is referred to as the remaining utility of $X_i$, and it represents the sum of the utilities of the remaining items such that \textit{sumRu}($X_i$) = $\sum_{T_j \subseteq \mathcal{D} \land X_i \subseteq T_j \wedge x_j \in T_j/X_i}$$u(x_j, T_j)$ \cite{liu2012mining}.

\textbf{Strategy 2}: If the sum of \textit{sumIu} and \textit{sumRu} in the utility-list of $X_i$ is higher than or equal to \textit{minUtil} ($\sigma$ and $\xi$ in this paper), the extension of $X_i$ and their transitive extensions are the potential HUIs. Otherwise, they are low-utility itemsets, and $X_i$ and its extensions can safely be ignored \cite{liu2012mining}.

It can be observed from Fig. \ref{fig:TP_tree_node} that the TP-tree becomes increasingly luxurious with increasing nodes. In fact, not all branches of the TP-tree contain target itemsets; therefore, TargetUM needs to cut off several useless branches in advance to avoid wasting construction and querying time. First, TargetUM needs to determine whether or not the current branch is worth using to locate the corresponding item using the first pruning strategy.

The TargetUM algorithm is intended to identify all HUIs that contain all target itemsets. When determining the target HUIs, it should be ensured that the search branches include the target patterns. Otherwise, the current querying operation is meaningless. In the TP-tree, each item of any path is sorted by \textit{TWU}-ascending order. Accordingly, the \textit{TWU} value of each node in the TP-tree is no less than that of its parent node. After sorting, the position of any node in the TP-tree can rapidly be located according to the item header table. In the following section, we describe how to filter out the useless branches.

\textbf{Strategy 3}: Select the item $p$ with \textit{TWU}$_{max}$ in the target itemsets, locate the position of node $p$ and current branch $\gamma$ in the TP-tree through the item header table, and query upwards to obtain the target utility pattern $X$. In TargetUM, $T^\prime$ is sorted by \textit{TWU}-ascending order and \textit{p.twu} $\ge$ \textit{p.parent.twu} according to \textbf{property 1}. Thus, for $i_\imath$, $i_\imath$ $\in$ $T^\prime$, if $i_\imath$ may exist in $\gamma$,  \textit{currentNode.twu} $\ge$ \textit{TWU}($i_\imath$). If \textit{currentNode.twu} $<$ \textit{TWU}($i_\imath$), $i_\imath$ does not exist in $\gamma$, and neither is $T^\prime$; thus, $\gamma$ is discarded. Note that more than one item may have the same \textit{TWU}; thus, if \textit{currentNode.twu} = \textit{TWU}($i_\imath$), and it is also necessary to determine whether \textit{currentNode.name} = $i_\imath$.

For example, assume an HUI = $\{$$d,a,g,e,c$$\}$ and \textit{$T^\prime$} = $\{$$e,c$$\}$. The \textit{TWU} of item $c$ is the largest and node $e$ is the parent node of $c$. In this case, $T^\prime$ is a subset of the HUI (\textit{c.twu} $\ge$\textit{e.twu}). If we set $T^\prime$ = $\{$$b,c$$\}$, the \textit{TWU} of $c$ is the largest. Subsequently, we need to determine \textit{b} and obtain \textit{e.twu}, \textit{g.twu}, and \textit{a.twu} $>$ \textit{TWU}($b$). Finally, we obtain \textit{d.twu} $<$ \textit{TWU}($b$), which means that $b$ is not in the HUI and neither is $T^\prime$ = \{$b, c$\}.

\subsection{Proposed targeted querying algorithm}

Based on the previous introduction, the proposed novel TargetUM algorithm can be described as follows:

\subsubsection{Target pattern trie construction}

The TP-tree plays a key role during the mining process. \textbf{Algorithm \ref{algo:construct_TP_tree}} demonstrates how to construct a TP-tree. It takes a HUI $X^\prime x$ ($X^\prime$ is initialized as null and $x$ is the last item of $X^\prime x$), and other information of $X^\prime$, such as \textit{IUs}, \textit{RUs}, \textit{TUs}, $u(x)$ (which refers to $\alpha$), and \textit{twu}$(x)$ (which refers to $\beta$) as input. A set of HUIs is obtained by HUI-Miner \cite{liu2012mining}. Once a new HUI is discovered, it is inserted into a TP-tree with target \textit{minUtil} $\sigma$. Lines 1 to 3 provide the three initial parameters. For each item $e \in X^\prime$, we can quickly determine the position (w.r.t. the index) of $e$ in the TP-tree (line 5). If $e$ does not exist, a new node $e$ is created and inserted into the TP-tree, following which \textbf{Algorithm \ref{algo:construct_item_header_table}} records the relative message of $e$ (lines 6 to 9). After processing all items of $X^\prime$, a complete TP-tree of $X^\prime$ is constructed (lines 4 to 13). Thereafter, in line 14, $x$ is determined by the binary search method. If no new node is identified, \textbf{Algorithm \ref{algo:construct_item_header_table}} is called (lines 15 to 17). In contrast, if $x$ is already stored in the TP-tree, it only updates certain relative information (lines 19 to 21). Finally, after dealing with the last item $x$, it will be the end of itemset $X^\prime x$, where $X^\prime x$ is a complete HUI (line 23).

\begin{algorithm}
	\caption{Construct TP-tree procedure}
	\label{algo:construct_TP_tree}
	\LinesNumbered
	\KwIn{$X^\prime$: the prefix of HUI; \textit{IUs}: the utility list of $X^\prime$; \textit{RUs}: the remaining utility list of $X^\prime$; \textit{TUs}: the \textit{twu} list of $X^\prime$; $x$: the last item of HUI; $\alpha$: the utility of $x$; $\beta$: the remaining utility of $x$; $\theta$: the \textit{twu} of $x$.}
	\KwOut{a TP-tree.}
	
	\textit{listNodes} $\leftarrow$ \textit{root.childs};
	
	\textit{currentNode} $\leftarrow$ \textit{NULL};
	
	\textit{parentNode} $\leftarrow$ \textit{NULL};
	
	\For{each element $e$ $\in$ $X^\prime$, \textit{e.iu} $\in$ \textit{IUs}, \textit{e.ru} and \textit{e.twu} $\in$ \textit{RUs}}{
		\textit{currentNode} $\leftarrow$ use binary search method to find $e$ in \textit{listNodes};
		
		\If{\textit{currentNode} == \textit{NULL}}{
			\textit{currentNode} $\leftarrow$ create a new node as \textbf{Node(\textit{e}, \textit{e.iu}, \textit{e.ru}, \textit{e.ru})};
			
			append \textbf{\textit{currentNode}} to \textbf{\textit{listNodes}};
			
			call \textbf{Item-Header Table}\textbf{(\textit{e}, \textit{currentNode})};
		}
		\textit{parentNode} $\leftarrow$ \textit{currentNode};
		
		\textit{listNodes} $\leftarrow$ \textit{currentNode.childs};
	}
	\textit{currentNode} $\leftarrow$ use binary search method to find $e$ in \textit{listNodes};
	
	\eIf{\textit{currentNode} == \textit{NULL}}{
		\textit{currentNode} $\leftarrow$ create a new node which \textbf{Node(\textit{x}, $\alpha$, $\beta$, $\theta$)};
		
		call \textbf{Item-Header Table}\textbf{(\textit{e}, \textit{currentNode})};
	}{
		\textit{currentNode.sumIu} $\leftarrow$ $\alpha$;
		
		\textit{currentNode.sumRu} $\leftarrow$ $\beta$;
		
		\textit{currentNode.twu} $\leftarrow$ $\theta$;
	}
	\textit{currentNode.isEnd} $\leftarrow$ \textit{TRUE}.
\end{algorithm}

\begin{algorithm}
	\caption{Item-Header Table procedure}
	\label{algo:construct_item_header_table}
	\LinesNumbered
	\KwIn{$x$: an item; \textit{node}: a node named $x$; \textit{mapItemNode}: store the header node of each item-header table; \textit{mapItemLastNode}: store the tail nodes of each item-header table.}
	\KwOut{an item-header table of $x$.}
	
	\textit{lastNode} $\leftarrow$ a node named $x$ $\in$ \textit{\textbf{mapItemLastNode}};
	
	\If{\textit{lastNode} != \textit{NULL}}{
		\textit{lastNode.link} $\leftarrow$ \textit{node};
	}
	
	append \textit{node} to \textit{\textbf{mapItemLastNode}} of $x$;
	
	\textit{headNode} $\leftarrow$ a node named $x$ $\in$ \textit{\textbf{mapItemNode}};
	
	\If{\textit{lastNode} == \textit{NULL}}{
		append \textit{node} to \textit{\textbf{mapItemLastNode}} of $x$;
	}
	\textbf{return} an item-header table of $x$;
\end{algorithm}

\subsubsection{Item header table construction}

The item header table plays a key role in the construction of a TP-tree. \textbf{Algorithm \ref{algo:construct_item_header_table}} lists the details regarding the construction procedure. The function of the item header table aims to build the node chains of interesting items. According to the mark header and tail indexes, with the binary searching method, the item header table makes it easy to obtain the real position of any item (lines 1 to 10). Finally, \textbf{Algorithm \ref{algo:construct_item_header_table}} returns an updated item header table containing $x$ (line 10).

\begin{algorithm}
	\caption{Output Prefix procedure}
	\label{algo:output_prefix}
	\LinesNumbered
	\KwIn{$T^\prime$: a target itemset; \textit{mapItemNode}: the header node of each item-header table; $\xi$: user-specified target \textit{minUtil}; \textit{mapItemToTwu}: a map of \textit{twu} values of each item.}
	\KwOut{\textit{THUIs}: a list of target high-utility itemsets.}
	
	\textit{currentNode} $\leftarrow$ \textit{NULL};
	
	\textit{posToMatch} $\leftarrow$ $|T^\prime|$;
	
	\textit{node} $\leftarrow$ the last item of $T^\prime$;
	
	\While{\textit{node} != \textit{NULL}}{
		\If{(\textit{node.sumIu} + \textit{node.sumRu}) $\ge$ $\xi$}{
			the current HUI $X$ $\leftarrow$ \textit{node.name};
			
			\textit{posToMatch} decrease 1;
			
			\textit{currentNode} $\leftarrow$ \textit{node.parent};
			
			\While{\textit{currentNode} != \textit{NULL}}{
				\If{\textit{posToMatch} $\ge$ 0}{
					a new item $y$ $\leftarrow$ $T^\prime$[posToMatch];
					
					\If{\textit{currentNode.twu} $<$ \textit{y.twu}}{
						\textbf{break};	
					}
					\If{\textit{currentNode.twu == \textit{Y.twu}} AND \textit{currentNode.item} == $y$}{
						\textit{posToMatch} decrease 1;
					}
				}
				$X$ $\leftarrow$ $X$ $\cup$ \textit{currentNode.name};
				
				\textit{currentNode} $\leftarrow$ \textit{currentNode.parent};
			}
			\If{posToMatch == -1}{
				\If{\textit{node.sumRu} $\ge$ $\xi$ AND \textit{node.isEnd} == \textit{TRUE}}{
					new \textit{THUIs} $\leftarrow$ \textit{THUIs} $\cup$ $X$;
				}
				call \textbf{Output Suffix}\textbf{(\textit{node.childs}, \textit{X})};	
			}
			\textit{node} $\leftarrow$ the next node of $T^\prime$;
		}
	}
\end{algorithm}

\begin{algorithm}
	\caption{Output Suffix procedure}
	\label{algo:output_suffix}
	\LinesNumbered
	\KwIn{\textit{Us}: a branch that containing target itemset; $X$: the prefix of a HUI; $\xi$: the target \textit{minUtil}.}
	\KwOut{\textit{THUIs}: a list of target high-utility itemsets.}
	
	\For{each \textit{node} $\in$ \textit{Us}}{
		$X^\prime$ $\leftarrow$ $X$ $\cup$ \textit{node.name};
		
		\If{\textit{node.sumIu} $\ge$ $\xi$ AND \textit{node.isEnd} == \textit{TRUE}}{
			update \textit{THUIs} $\leftarrow$ \textit{THUIs} $\cup$ $X^\prime$;
		}
		\If{\textit{node.childs} != \textit{NULL} AND \textit{node.sumIu} + \textit{node.sumRu} $\ge$ $\xi$}{
			call \textbf{outputSuffix}\textbf{(\textit{node.childs}, \textit{X$^\prime$})};
		}
	}
\end{algorithm}

\subsubsection{Output function}

All of the target HUIs can be discovered after constructing the complete TP-tree. As any THUI must be an HUI ($\forall$THUI $\subseteq$ HUIs), several situations may exist: a) the THUI is located at the end of the HUI; b) the THUI is located in the primary of the HUI; c) the THUI is located in the HUI; d) the THUI continuously occurs in the HUI; and e) the THUI discontinuously occurs in the HUI. To this end, we design two methods to deal with these five situations. To ensure the correctness and completeness of the discovered results, we adopt the depth-first searching method. The details are provided below:

We first determine in advance whether or not a prefix of the THUI exists (\textbf{Algorithm \ref{algo:output_prefix}}). The last item of the target itemset $T^\prime$ is set as the \textit{node}. According to \textit{node} and \textit{mapItemNode}, we can easily obtain all of the information regarding the current HUI (a branch includes \textit{node}). The \textit{posToMatch} parameter is used to record the current two items (lines 1 to 3). Subsequently, we traverse each item of $T^\prime$. As demonstrated in \textbf{strategy 2}, if \textit{sumIu} + \textit{sunRu} is no less than $\xi$, the current itemset $X$ may be a THUI, and the remaining items should be compared further (lines 4 to 30). However, in line 12, if the \textit{TWU} of \textit{currentNode} is less than the current target item $y$, the comparison process can be directly aborted (\textbf{strategy 3}). Once the \textit{posToMatch} is equal to -1, we have identified an entire target itemset, where $T^\prime \subseteq X$, following which TargetUM can call \textbf{Algorithm \ref{algo:output_suffix}} to output this new THUI (lines 22 to 27).

\textbf{Algorithm \ref{algo:output_suffix}} is more succinct than \textbf{Algorithm \ref{algo:output_prefix}}. It is a recursive method for exploring all of the suffix nodes of $X$. The expanded HUI $ X ^ \prime $ can be obtained for each \textit{node} $\in$ \textit{Us} (line 2). Subsequently, TargetUM filters out non-target HUIs and stores the THUIs using a list (lines 3 to 4). Finally, according to \textbf{strategy 2}, it determines whether to continue exploring the suffix nodes (lines 6 to 7).

\section{Performance Evaluation}
\label{sec:experiments}

To the best of our knowledge, THUIM has not yet been studied. Therefore, there is no suitable algorithm for comparison. To determine the effectiveness of our novel algorithm, we conducted an extensive experiment to compare TargetUM with the HUI-Miner algorithm \cite{liu2012mining} directly. The function of HUI-Miner in the experiments was executed as target post-processing. This can be regarded as a benchmark experiment to verify the accuracy of the results discovered by TargetUM. Both real-life and synthetic datasets were used to verify the effectiveness of TargetUM.

\subsection{Target post-processing}
\label{sec:target_post_processing}

In short, only simple modifications were made to the original HUI-Miner algorithm, and the user-defined target patterns were extracted from the HUIs using a post-processing mechanism. The HUIs that were discovered by HUI-Miner were checked by \textit{T'}. If each item in \textit{$T^\prime$} was included in the HUIs, this HUI could be said to be a THUI. For example, given two HUIs, $\{d, a, g, e\}$ and $\{d, a, g, e, c\}$, and the user-defined target pattern string was $\{e, c\}$. Following post-processing, $\{d, a, g, e, c\}$ contained the target pattern $\{e, c\}$; thus, it was a THUI. However, $\{d, a, g, e\}$ could not be a THUI because it did not contain item $c$ in \textit{$T^\prime$}. Note that this post-processing was used as a verification algorithm to verify the accuracy of the results of TargetUM. The HUI-Miner algorithm with a post-processing mechanism is denoted as HUI-mp.

\subsection{Experimental setup and dataset}

Experiments were conducted on four real-life datasets, as shown in Table \ref{tab:dataset}. The characteristics of these datasets include: 1) the name of the dataset; 2) the size of each dataset; 3) the number of transactions for each dataset; 4) the number of items for each dataset; 5) the average length (AvgLen) of the transaction; and 6) the maximum length (MaxLen) of the transaction. Both TargetUM and HUI-mp were implemented using the Java language and executed on a PC with an AMD Ryzen 5 3600 6-Core Processor 3.60 GHz and 8 GB of memory, running on a 64-bit Microsoft Windows 10 platform.

\begin{table}[!htb]
	\centering	
	\caption {Characteristics of different datasets \label{tab:dataset}}
	\begin{tabular}{|l|c|c|c|c|c|}
		\hline 
		\textbf{Dataset} & \textbf{Size (KB)} & \textbf{Trans} & \textbf{Items} & \textbf{AvgLen} & \textbf{MaxLen} \\  \hline
		chainstore & 81102 & 1112950 & 46086 & 7.2 & 170 \\
		ecommerce & 1968 & 14976 & 3468 & 11.6 & 29 \\
		retail & 1845 & 24735 & 11371 & 10.3 & 74 \\
		foodmart & 176 & 4141 & 1559 & 4.4 & 14 \\ \hline 	
	\end{tabular}	
\end{table}

Foodmart, chainstore, and ecommerce are three real-life customer transaction datasets with real utility values. The foodmart dataset originates from a retail store, and it was obtained and transformed from SQL-Server 2000, with 4,141 transactions and 1,559 items. The chainstore dataset was obtained from a major grocery store chain in California, USA, and it was transformed from the NU-Mine Bench, with 1,112,949 transactions and 46,086 items. The ecommerce dataset contains 14,975 transactions and 3,486 items. Retail is a real dataset with synthetic utility values, and the internal utility values are generated using a uniform distribution in [1, 10]. Thus, retail consists mainly of customer transactions from an anonymous Belgian retail store.

In the following experiments, TargetUM was compared with several variants based on the use of different pruning strategies. These are denoted as \textbf{TargetUM$_{13}$} (with pruning strategies 1 and 3), \textbf{TargetUM}$_{23}$ (with pruning strategies 2 and 3), \textbf{TargetUM}$_3$ (with pruning strategy 3), and \textbf{TargetUM} (with all pruning strategies), respectively. These four versions were used to evaluate the efficiency of TargetUM.

\subsection{Itemset analysis}

\begin{table}[htb]
	\fontsize{5pt}{9pt}\selectfont
	\centering
	\caption{Derived itemsets under varied $\sigma$ ($\sigma$ = $\xi$)}
	\label{table:thresholdsigma}
	\begin{tabular}{c|c|llllll}
		\hline \hline
		\multirow{2}{*}{\textbf{Dataset}} & \multirow{2}{*}{\textbf{Pattern}} & \multicolumn{6}{c}{\textbf{\# itemsets by varying threshold $\sigma$ ($\sigma$ = $\xi$)}} \\ \cline{3-8}
		& &$\sigma_1$ & $\sigma_2$ & $\sigma_3$ & $\sigma_4$ & $\sigma_5$ & $\sigma_6$ \\ \hline
		chainstore & THUIs* & 267 & 207 & 170 & 131 & 106 & 91 \\
		$\{16967\}$ & THUIs & 267 & 207 & 170 & 131 & 106 & 91 \\ \hline
		ecommerce & THUIs* & 4,643,198 & 2,463,092 & 1,641,208 & 1,050,396 & 596,385 & 293,757 \\
		$\{150561222\}$ & THUIs & 4,643,198 & 2,463,092 & 1,641,208 & 1,050,396 & 596,385 & 293,757 \\ \hline
		retail & THUIs* &34,240  & 26 & 19 & 16 & 14 & 12 \\
		$\{976\}$ & THUIs & 34,240 & 26 & 19 & 16 & 14 & 12 \\ \hline
		foodmart & THUIs* & 9,214 & 9,196 & 8,833 & 7,727 & 5,896 & 3,494 \\
		$\{1340\}$ & THUIs & 9,214 & 9,196 & 8,833 & 7,727 & 5,896 & 3,494 \\ \hline 
		\hline
	\end{tabular}
\end{table}

\begin{table}[htb]
	\fontsize{5pt}{9pt}\selectfont
	\centering
	\caption{Derived itemsets under varied $T^\prime$ ($\sigma$ = $\xi$)}
	\label{table:thresholdT}
	\begin{tabular}{c|c|llllll}
		\hline \hline
		\multirow{2}{*}{\textbf{Dataset}} & \multirow{2}{*}{\textbf{Pattern}} & \multicolumn{6}{c}{\textbf{\# itemsets by varying $T^\prime$}} \\ \cline{3-8}
		& &$T^\prime_1$ & $T^\prime_2$ & $T^\prime_3$ & $T^\prime_4$ & $T^\prime_5$ & $T^\prime_6$ \\ \hline
		chainstore & THUIs* & 267 & 20 & 16 & 100 & 3 & 3 \\
		$\sigma$ = 300000 & THUIs & 267 & 20 & 16 & 100 & 3 & 3 \\ \hline
		ecommerce & THUIs* & 3,250,046 & 2,200,992 & 811,970 & 506,036 & 390,088 & 240,666 \\
		$\sigma$ = 550000 & THUIs & 3,250,046 & 2,200,992 & 811,970 & 506,036 & 390,088 & 240,666 \\ \hline
		retail & THUIs* & 34,213 & 34,200 & 34,200 & 34,079 & 34,065 & 34,019 \\
		$\sigma$ = 1370 &THUIs & 34,213 & 34,200 & 34,200 & 34,079 & 34,065 & 34,019 \\ \hline
		foodmart & THUIs* & 4,096 & 64 & 32 & 32 & 4 & 416 \\
		$\sigma$ = 20 & THUIs & 4,096 & 64 & 32 & 32 & 4 & 416 \\ \hline 
		\hline
		
	\end{tabular}
\end{table}

\begin{table}[htb]
	\fontsize{5pt}{9pt}\selectfont
	\centering
	\caption{Derived itemsets under varied $\xi$}
	\label{table:thresholdxi}
	\begin{tabular}{c|c|llllll}
		\hline \hline
		\multirow{2}{*}{\textbf{Dataset}} & \multirow{2}{*}{\textbf{Pattern}} & \multicolumn{6}{c}{\textbf{\# itemsets by varying threshold $\xi$}} \\ \cline{3-8}
		& &$\xi_1$ & $\xi_2$ & $\xi_3$ & $\xi_4$ & $\xi_5$ & $\xi$ \\ \hline
		\multirow{2}{*}{chainstore} & THUIs* & 267 & 207 & 170 & 131 & 106 & 91 \\
		& THUIs & 267 & 207 & 170 & 131 & 106 & 91 \\ \hline
		\multirow{2}{*}{ecommerce} & THUIs* & 4,643,198 & 2,463,092 & 1,641,208 & 1,050,396 & 596,385 & 293,757 \\
		& THUIs & 4,643,198 & 2,463,092 & 1,641,208 & 1,050,396 & 596,385 & 293,757 \\ \hline
		\multirow{2}{*}{retail} & THUIs* &34,240  & 26 & 19 & 16 & 14 & 12 \\
		& THUIs & 34,240 & 26 & 19 & 16 & 14 & 12 \\ \hline
		\multirow{2}{*}{foodmart} & THUIs* & 9,214 & 9,196 & 8,833 & 7,727 & 5,896 & 3,494 \\
		& THUIs & 9,214 & 9,196 & 8,833 & 7,727 & 5,896 & 3,494 \\ \hline 
		\hline
	\end{tabular}
\end{table}

TargetUM was initially proposed to solve the target HUM problem. The main goal is to obtain accurate targeted HUPs. Thus, the results were verified in the form of a comparison with HUI-mp. HUI-mp used post-processing to verify the results. Note that, for comparison purposes, the target HUIs obtained by TargetUM are denoted as THUIs*, whereas those obtained by HUI-mp are denoted as THUIs. By analyzing the datasets, we set different thresholds and $T^\prime$ for different datasets.

We compared the number of itemsets on the test datasets to demonstrate the effect of varying the parameters $\sigma$, $\xi$, and $T^\prime$; the results for various parameters are displayed in Tables \ref{table:thresholdsigma}, \ref{table:thresholdT}, and \ref{table:thresholdxi}, respectively. As can be observed from these tables, THUIs* and THUIs obtained the same count. For Table \ref{table:thresholdsigma}, letting $\sigma$ = $\xi$, and $\sigma$ or $\xi$ remain unchanged, with the increase in $\sigma$ ($\xi$), the count of the THUIs became increasingly smaller. For Table \ref{table:thresholdT}, with $T^\prime$ as a parameter, with either a single item or an itemset, multiple $T^\prime$ were set for each dataset. Note that no direct relationship existed between $T^\prime_k$ and $T^\prime_{k+1}$ because of the limitation of the average transaction length of the dataset, such as chainstore and foodmart.

Table \ref{table:thresholdxi} satisfies multiple target queries under the same minimum utility threshold with varying $\xi$, fixed $\sigma$, and fixed $T^\prime$. We set the chainstore $\sigma$ = 300000, $T^\prime$ = $\{16967\}$, the ecommerce $\sigma$ = 550000, $T^\prime$ = $\{150561222\}$, the retail $\sigma$ = 1370, $T^\prime$ = $\{976\}$, and the foodmart $\sigma$ = 20, $T^\prime$ = $\{1340\}$. We can obtain the same results from Table \ref{table:thresholdxi} as those in Table \ref{table:thresholdsigma}. According to the above analysis, it can be concluded that the proposed TargetUM algorithm is effective for the THUIM problem.

\subsection{Efficiency analysis}

\begin{figure}[!htb]
	\centering
	\includegraphics[height=0.36\textheight,width=1\linewidth,trim=60 0 50 0,clip,scale=0.45]{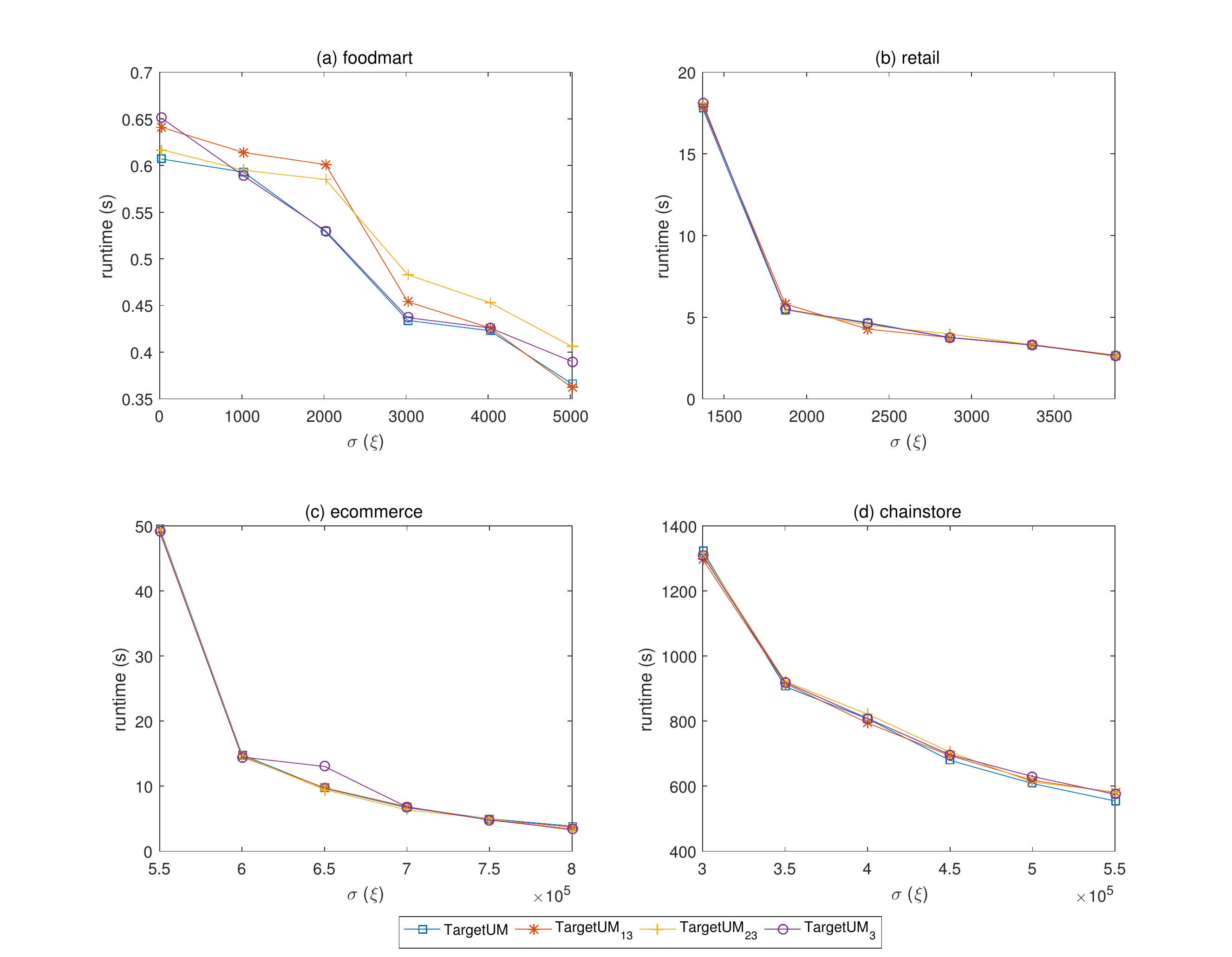}
	\caption{Runtime under varied $\sigma$ and $\xi$ with $\sigma$ = $\xi$ and a fixed $T^\prime$. (a) foodmart ($\sigma$ = 20, $T^\prime$ = $\{1340\}$). (b) retail ($\sigma$ = 1370, $T^\prime$ = $\{976\}$). (c) ecommerce ($\sigma$ = 550000, $T^\prime$ = $\{150561222\}$). (d) chainstore ($\sigma$ = 300000, $T^\prime$ = $\{16967\}$)}
	\label{fig:runtime}
\end{figure}

\begin{figure}[!htb]
	\centering
	\includegraphics[height=0.36\textheight,width=1\linewidth,trim=60 0 50 0,clip,scale=0.45]{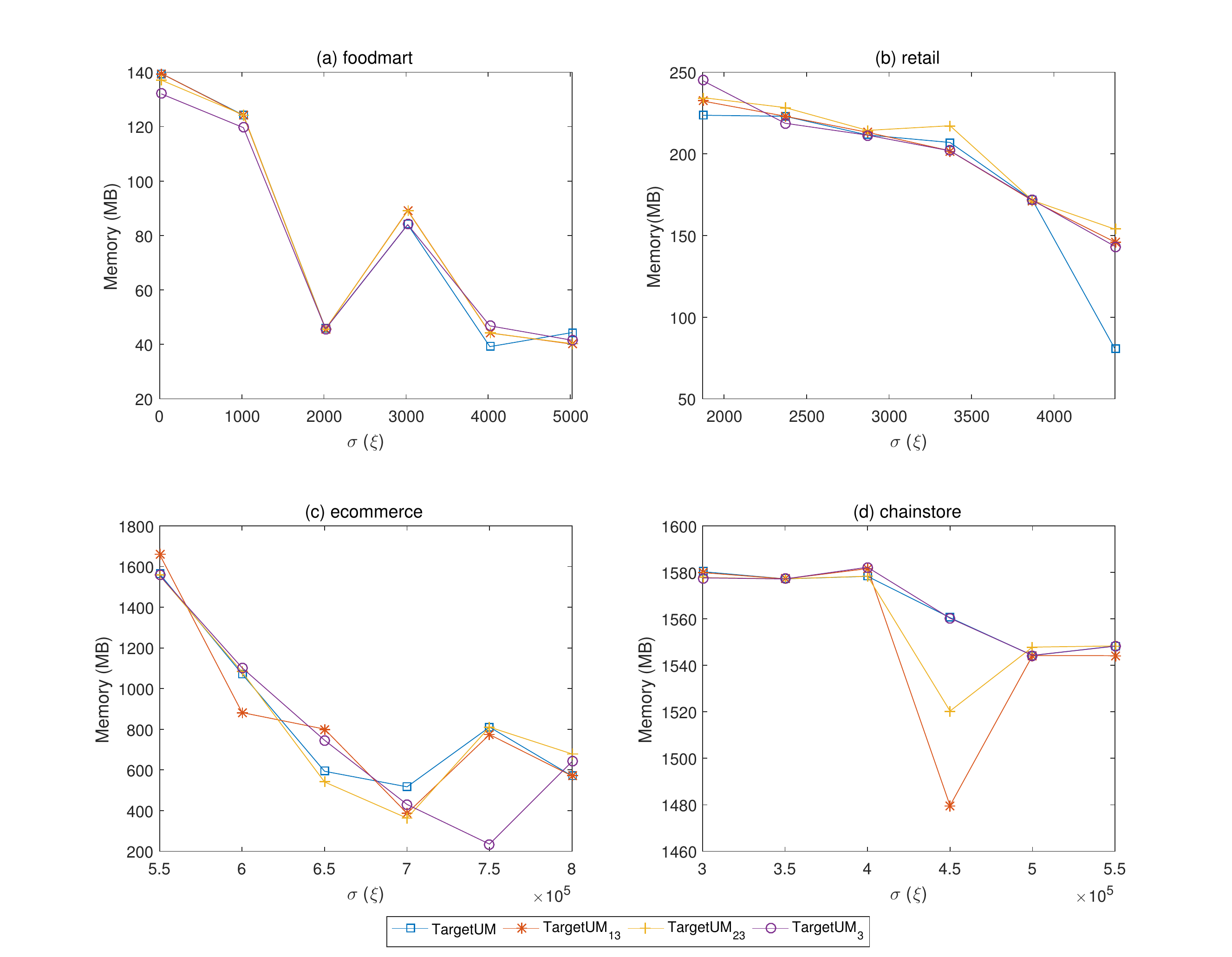}
	\caption{Memory under varied $\sigma$ and $\xi$ with $\sigma$ = $\xi$ and a fixed $T^\prime$. (a) foodmart ($\sigma$ = 20, $T^\prime$ = $\{1340\}$). (b) retail ($\sigma$ = 1370, $T^\prime$ = $\{976\}$). (c) ecommerce ($\sigma$ = 550000, $T^\prime$ = $\{150561222\}$). (d) chainstore ($\sigma$ = 300000, $T^\prime$ = $\{16967\}$)}
	\label{fig:memory}
\end{figure}

\begin{figure}[!h]
	\centering
	\includegraphics[height=0.36\textheight,width=1.02\linewidth,trim=60 0 50 0,clip,scale=0.45]{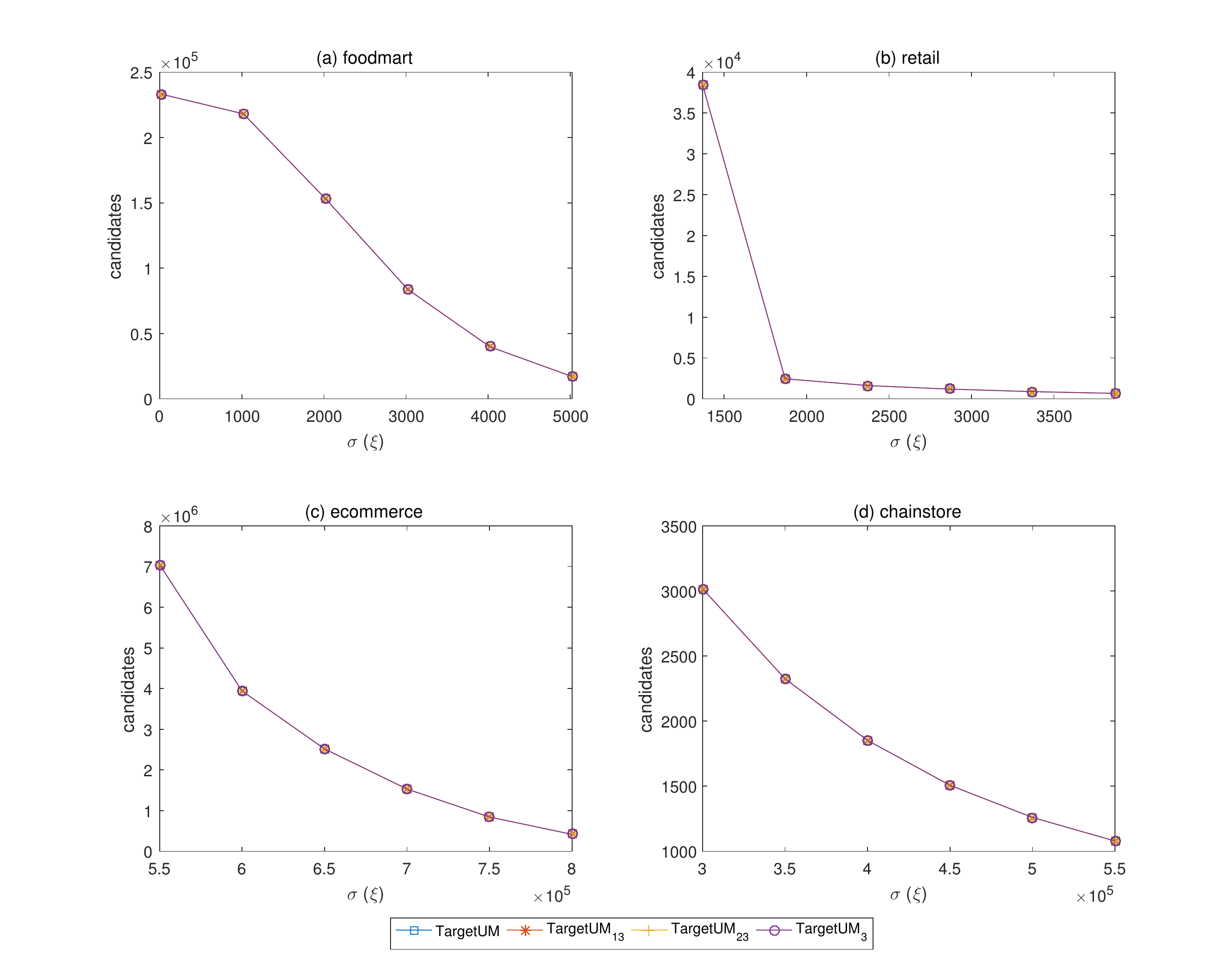}
	\caption{Candidates of the compared methods under varied $\sigma$ and $\xi$ with $\sigma$ = $\xi$ and a fixed $T^\prime$. (a) foodmart ($\sigma$ = 20, $T^\prime$ = $\{1340\}$). (b) retail ($\sigma$ = 1370, $T^\prime$ = $\{976\}$). (c) ecommerce ($\sigma$ = 550000, $T^\prime$ = $\{150561222\}$). (d) chainstore ($\sigma$ = 300000, $T^\prime$ = $\{16967\}$)}
	\label{fig:candities}
\end{figure}

To evaluate the overall performance of the proposed algorithm more effectively, the following analysis was performed in terms of the runtime, memory consumption, and candidates. The detailed results are presented in Figs. \ref{fig:runtime} to \ref{fig:candities}.

Fig. \ref{fig:runtime} depicts the running time of four variants of the TargetUM algorithm. We set varying $\sigma$ and $\xi$ with $\sigma$ = $\xi$ and a fixed $T^\prime$. As can be observed in Fig. \ref{fig:runtime}, the overall running time gradually decreased with an increase in $\sigma$ or $\xi$. The TargetUM with the three strategies had the shortest running time. Strategies 1 and 2 could prune most of the search space. It is interesting to note that the overall running time of the four versions of TargetUM did not differ significantly, owing to the limitation of the TP-tree, which is a HUI-based tree and also a set-enumeration tree. Therefore, limited by the actual storage, the number of HUIs for building the TP-tree could be too high, which led to the algorithm searching the end of the tree very quickly, and only a slight difference can be observed from Fig. \ref{fig:runtime}. This is clear from Fig. \ref{fig:runtime}(a), in which foodmart was a sparse dataset and the overall running time that was required was very short; thus, the difference could be expressed. However, for the retail, ecommerce, and chainstore datasets, the difference was not easy to express.

Figs. \ref{fig:memory} and \ref{fig:candities} present the memory usage and number of candidate itemsets of the algorithm. It can be observed from the figures that the memory usage of the four versions of the TargetUM algorithm as a whole gradually decreased with the increase in the threshold value and tended to be consistent, because TargetUM first constructed the TP-tree using HUIs and then determined the THUIs that contained $T^\prime$ by searching the TP-tree. Therefore, the search process did not affect the memory usage. Similarly, as any HUI was likely to be a THUI, the HUIs were candidate itemsets and were fixed.

\subsection{Effect of pruning strategies}

\begin{figure}[!h]
	\centering
	\includegraphics[height=0.36\textheight,width=1.02\linewidth,trim=60 10 50 0,clip,scale=0.46]{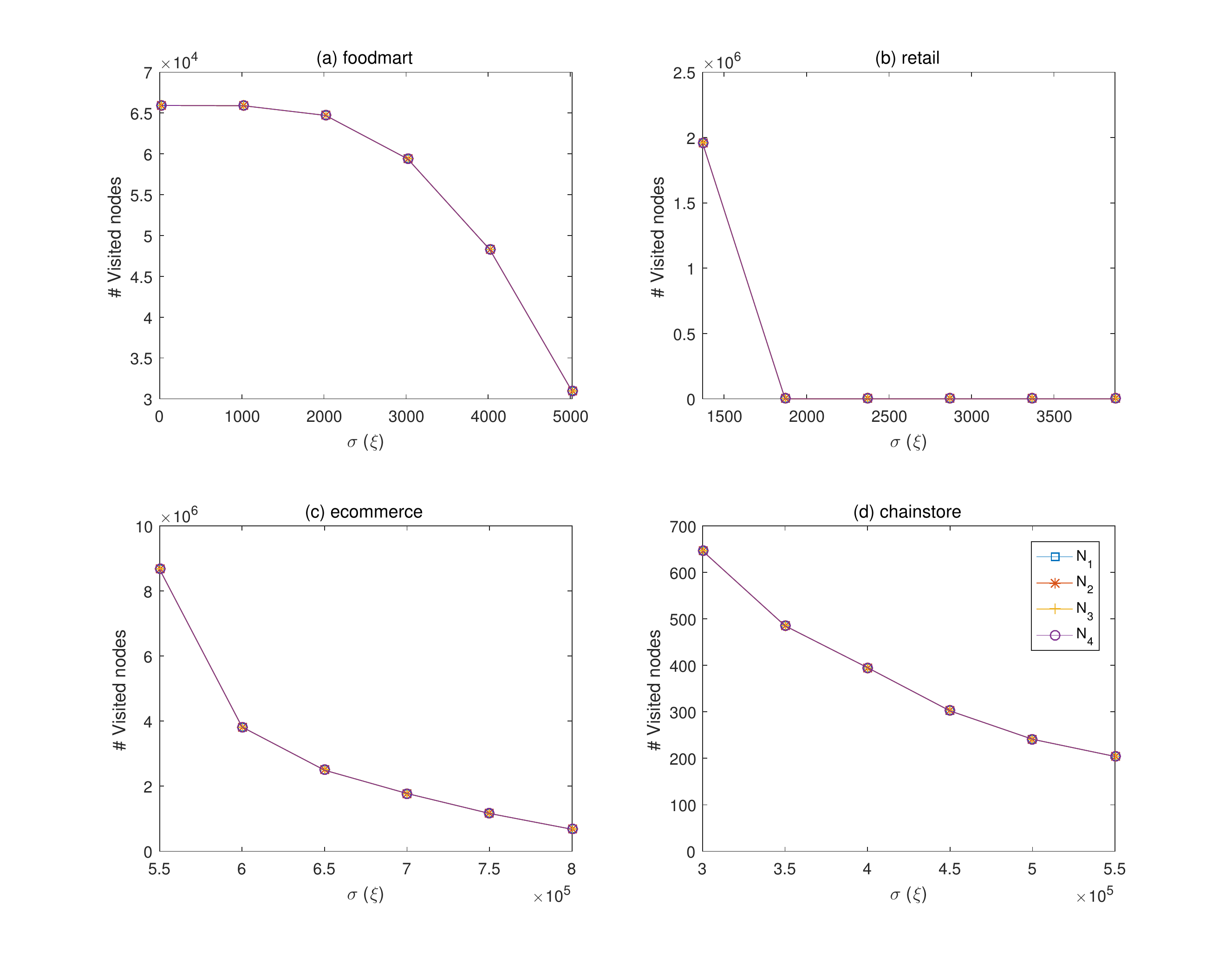}
	\caption{Number of visited nodes under varied $\sigma$ and $\xi$ with a fixed $T^\prime$. (a) foodmart ($\sigma$ = $\xi$ = 20, $T^\prime$ = $\{1340\}$). (b) retail ($\sigma$ = $\xi$ = 1370, $T^\prime$ = $\{976\}$). (c) ecommerce ($\sigma$ = $\xi$ = 550000, $T^\prime$ = $\{150561222\}$). (d) chainstore ($\sigma$ = $\xi$ = 300000, $T^\prime$ = $\{16967\}$)}
	\label{fig:nodefixed}
\end{figure}

\begin{figure}[!h]
	\centering
	\includegraphics[height=0.35\textheight,width=1.02\linewidth,trim=60 5 50 0,clip,scale=0.48]{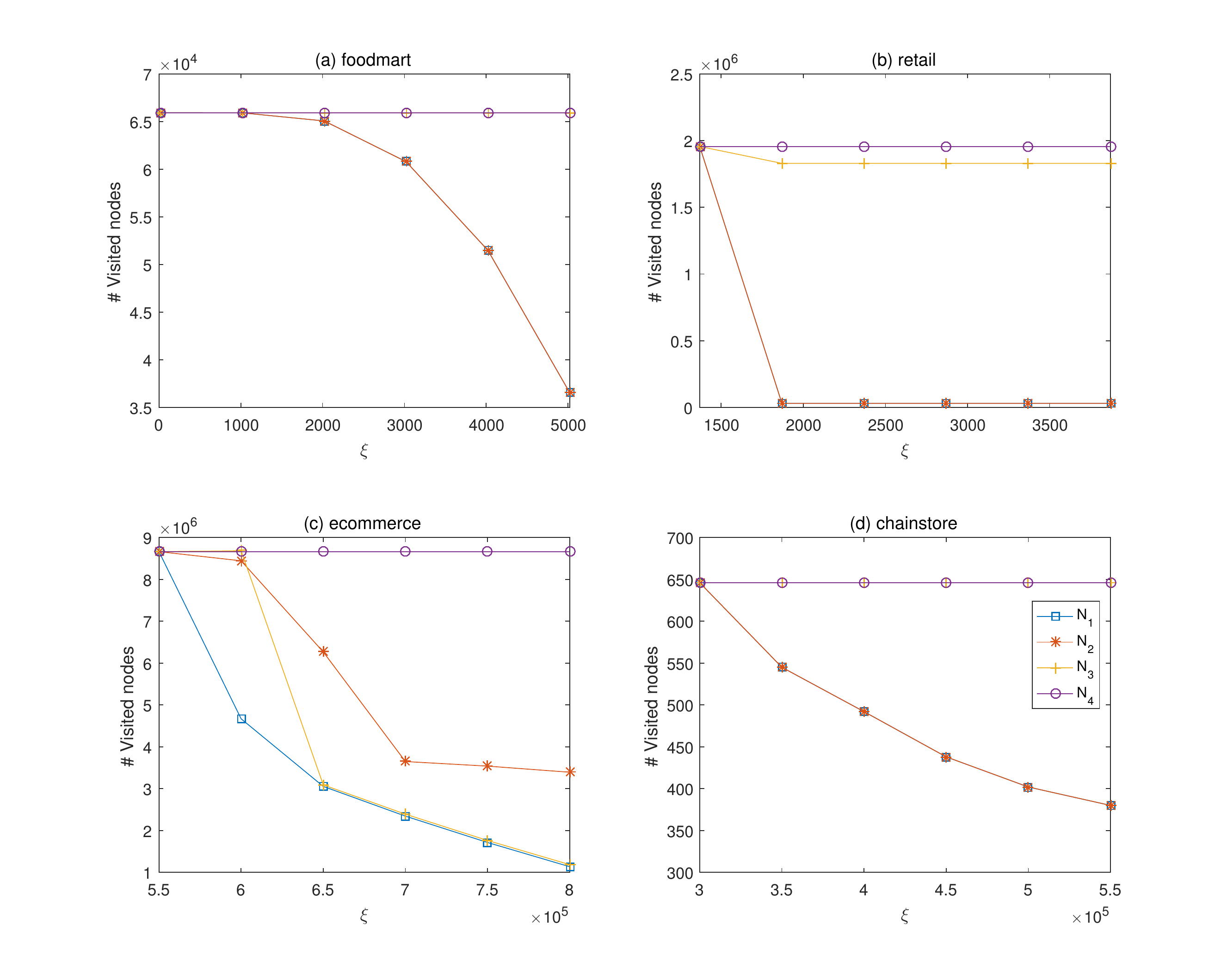}
	\caption{Number of visited nodes under varied $\xi$ with a fixed $\sigma$ and a fixed $T^\prime$. (a) foodmart ($\sigma$ = 20, $T^\prime$ = $\{1340\}$). (b) retail ($\sigma$ = 1370, $T^\prime$ = $\{976\}$). (c) ecommerce ($\sigma$ = 550000, $T^\prime$ = $\{150561222\}$). (d) chainstore ($\sigma$ = 300000, $T^\prime$ = $\{16967\}$)}
	\label{fig:nodestrategy}
\end{figure}

To evaluate the pruning strategy further, the number of visited nodes in the TP-tree was compared. For comparison purposes, the number of visited nodes of TargetUM (with all pruning strategies) is denoted as $N_1$, TargetUM$_{13}$ is denoted as $N_2$, TargetUM$_{23}$ is denoted as $N_3$, and TargetUM$_3$ is denoted as $N_4$. Figs. \ref{fig:nodefixed} and \ref{fig:nodestrategy} present the experimental results with different parameter settings.

Fig. \ref{fig:nodefixed} indicates a varying $\sigma$ = $\xi$ with a fixed $T^\prime$, and $\sigma$ and $\xi$ were changed accordingly. In Fig. \ref{fig:nodestrategy}, $\xi$ was varied with a fixed $\sigma$ and a fixed $T^\prime$. The following information can be obtained based on Figs. \ref{fig:nodefixed} and \ref{fig:nodestrategy}:

\begin{enumerate}[]
	\item Given $\sigma$ = $\xi$, equivalent to a regular query, all HUIs that contain $T^\prime$ must be the interesting THUIs. The pruning strategies were not very useful; thus, it can be observed that $N_1$ = $N_2$ = $N_3$ = $N_4$. In general, the number of visited nodes decreased as $\sigma$ or $\xi$ increased.
	
	\item The TargetUM algorithm first constructed the TP-tree and finally processed the query on the tree. The pruning strategies mainly worked on the TP-tree, as illustrated in Fig. \ref{fig:nodestrategy}. When $\xi$ was set to be varying with a fixed $\sigma$, and then multiple queries were performed, the number of visited nodes decreased as $\xi$ increased.
	
	\item Comparing $N_1$ and $N_4$, it can be observed that $N_4$ was always the same, because the entire TP-tree was searched on the fly. Strategy 3 was only dependent on an order, and the $T^\prime$ set in the experiment had only one item. For $T^\prime$, which had $|T^\prime|$ $>$ 1, strategy 3 worked. Compared to $N_4$, $N_1$ pruned a large part of the search space in the TP-tree. Thus, TargetUM$_1$ always consumed the shortest runtime among TargetUM (with all pruning strategies), TargetUM$_{13}$, TargetUM$_{23}$, and TargetUM$_3$, as indicated in Fig. \ref{fig:runtime}.
	
	\item Comparing $N_2$ and $N_3$ from Fig. \ref{fig:nodestrategy}(b) and \ref{fig:nodestrategy}(c), it can be concluded that both strategies 1 and 2 had a positive effect on accelerating the queries, but the results exhibited differences. Strategy 1 acted on the prefix and strategy 2 acted on the suffix, both of which played a very important role. Finally, strategies 1 and 2 both had a positive effect on reducing the search space, but exhibited variability with the sparsity of the dataset and $|T^\prime|$.
	
	\item $T^\prime$ is a significant factor for improving the mining performance on the THUI querying task.
\end{enumerate}

\section{Conclusions}
\label{sec:conclusion}

Existing utility-driven mining algorithms can only determine all HUIs that satisfy $\sigma$ in the transaction database, and not all of these itemsets are interesting. In this study, we first proposed the target utility mining problem and designed an algorithm known as TargetUM to handle this task. Target pattern mining always occupies substantial memory when stored, but storage resources are limited. A utility-based trie tree, namely a TP-tree, was designed. TargetUM adopts two utility threshold constraints to obtain the desired results and can also save on the memory cost. TargetUM can discover THUIs under the constraints of $T^\prime$, $\sigma$, and $\xi$. Moreover, conducted experiments on real and synthetic datasets to demonstrate that the TargetUM algorithm can obtain complete and accurate target results. 

It is a significant challenge to reduce the execution time and memory consumption in the addressed task. TargetUM first calculates all HUIs, then constructs the TP-tree, and finally determines the THUIs. Therefore, in the future, we intend to study more efficient algorithms and new data structures that can save on the execution time and memory cost while improving the performance. In particular, extending the algorithm to process big data in distributed systems will be a quite challenging task.

\ifCLASSOPTIONcaptionsoff
  \newpage
\fi

\bibliographystyle{IEEEtran}
\bibliography{TargetUM}

\begin{thebibliography}{10}
\providecommand{\url}[1]{#1}
\csname url@samestyle\endcsname
\providecommand{\newblock}{\relax}
\providecommand{\bibinfo}[2]{#2}
\providecommand{\BIBentrySTDinterwordspacing}{\spaceskip=0pt\relax}
\providecommand{\BIBentryALTinterwordstretchfactor}{4}
\providecommand{\BIBentryALTinterwordspacing}{\spaceskip=\fontdimen2\font plus
\BIBentryALTinterwordstretchfactor\fontdimen3\font minus
  \fontdimen4\font\relax}
\providecommand{\BIBforeignlanguage}[2]{{%
\expandafter\ifx\csname l@#1\endcsname\relax
\typeout{** WARNING: IEEEtran.bst: No hyphenation pattern has been}%
\typeout{** loaded for the language `#1'. Using the pattern for}%
\typeout{** the default language instead.}%
\else
\language=\csname l@#1\endcsname
\fi
#2}}
\providecommand{\BIBdecl}{\relax}
\BIBdecl

\bibitem{mccarthy1989architecture}
D.~McCarthy and U.~Dayal, ``The architecture of an active database management
  system,'' \emph{ACM SIGMOD Record}, vol.~18, no.~2, pp. 215--224, 1989.

\bibitem{stonebraker1991postgres}
M.~Stonebraker and G.~Kemnitz, ``The postgres next generation database
  management system,'' \emph{Communications of the ACM}, vol.~34, no.~10, pp.
  78--92, 1991.

\bibitem{aggarwal2014frequent}
C.~C. Aggarwal, M.~A. Bhuiyan, and M.~Al~Hasan, ``Frequent pattern mining
  algorithms: A survey,'' \emph{Frequent Pattern Mining}, p.~19, 2014.

\bibitem{agrawal1994fast}
R.~Agrawal and R.~Srikant, ``Fast algorithms for mining association rules,'' in
  \emph{Proceedings of the 20-th ACM International Conference on Very Large
  Data Bases}, vol. 1215.\hskip 1em plus 0.5em minus 0.4em\relax Citeseer,
  1994, pp. 487--499.

\bibitem{han2000mining}
J.~Han, J.~Pei, and Y.~Yin, ``Mining frequent patterns without candidate
  generation,'' \emph{ACM SIGMOD Record}, vol.~29, no.~2, pp. 1--12, 2000.

\bibitem{fournier2017survey}
P.~Fournier-Viger, J.~C.~W. Lin, R.~U. Kiran, Y.~S. Koh, and R.~Thomas, ``A
  survey of sequential pattern mining,'' \emph{Data Science and Pattern
  Recognition}, vol.~1, no.~1, pp. 54--77, 2017.

\bibitem{gan2017data}
W.~Gan, J.~C.~W. Lin, H.~C. Chao, and J.~Zhan, ``Data mining in distributed
  environment: a survey,'' \emph{Wiley Interdisciplinary Reviews: Data Mining
  and Knowledge Discovery}, vol.~7, no.~6, p. e1216, 2017.

\bibitem{gan2021survey}
W.~Gan, J.~C.~W. Lin, P.~Fournier~Viger, H.~C. Chao, V.~Tseng, and P.~S. Yu,
  ``A survey of utility-oriented pattern mining,'' \emph{IEEE Transactions on
  Knowledge and Data Engineering}, vol.~33, no.~4, pp. 1306--1327, 2021.

\bibitem{liu2012mining}
M.~Liu and J.~Qu, ``Mining high utility itemsets without candidate
  generation,'' in \emph{Proceedings of the 21-st ACM International Conference
  on Information and Knowledge Management}, 2012, pp. 55--64.

\bibitem{fournier2014fhm}
P.~Fournier~Viger, C.~Wu, S.~Zida, and V.~S. Tseng, ``{FHM}: Faster
  high-utility itemset mining using estimated utility co-occurrence pruning,''
  \emph{Foundations of Intelligent Systems}, pp. 83--92, 2014.

\bibitem{lin2016fhn}
J.~C. Lin, P.~Fournierviger, and W.~Gan, ``{FHN}: An efficient algorithm for
  mining high-utility itemsets with negative unit profits,'' \emph{Knowledge
  Based Systems}, vol. 111, pp. 283--298, 2016.

\bibitem{ahmed2009efficient}
C.~F. Ahmed, S.~K. Tanbeer, B.~S. Jeong, and Y.~K. Lee, ``Efficient tree
  structures for high utility pattern mining in incremental databases,''
  \emph{IEEE Transactions on Knowledge and Data Engineering}, vol.~21, no.~12,
  pp. 1708--1721, 2009.

\bibitem{tseng2012efficient}
V.~S. Tseng, B.~E. Shie, C.~W. Wu, and P.~S. Yu, ``Efficient algorithms for
  mining high utility itemsets from transactional databases,'' \emph{IEEE
  Transactions on Knowledge and Data Engineering}, vol.~25, no.~8, pp.
  1772--1786, 2012.

\bibitem{yun2014high}
U.~Yun, H.~Ryang, and K.~H. Ryu, ``High utility itemset mining with techniques
  for reducing overestimated utilities and pruning candidates,'' \emph{Expert
  Systems with Applications}, vol.~41, no.~8, pp. 3861--3878, 2014.

\bibitem{zida2017efim}
S.~Zida, P.~Fournier~Viger, J.~C.~W. Lin, C.~Wu, and V.~S. Tseng, ``{EFIM}: a
  fast and memory efficient algorithm for high-utility itemset mining,''
  \emph{Knowledge and Information Systems}, vol.~51, no.~2, pp. 595--625, 2017.

\bibitem{liu2005two}
Y.~Liu, W.~Liao, and A.~Choudhary, ``A two-phase algorithm for fast discovery
  of high utility itemsets,'' in \emph{Pacific Asia Conference on Knowledge
  Discovery and Data Mining}.\hskip 1em plus 0.5em minus 0.4em\relax Springer,
  2005, pp. 689--695.

\bibitem{kubat2003itemset}
M.~Kubat, A.~Hafez, V.~V. Raghavan, J.~R. Lekkala, and W.~K. Chen, ``Itemset
  trees for targeted association querying,'' \emph{IEEE Transactions on
  Knowledge and Data Engineering}, vol.~15, no.~6, pp. 1522--1534, 2003.

\bibitem{shabtay2018guided}
L.~Shabtay, R.~Yaari, and I.~Dattner, ``A guided {FP-G}rowth algorithm for
  multitude-targeted mining of big data,'' \emph{arXiv preprint
  arXiv:1803.06632}, 2018.

\bibitem{abeysinghe2017query}
R.~Abeysinghe and L.~Cui, ``Query-constraint-based association rule mining from
  diverse clinical datasets in the national sleep research resource,'' in
  \emph{IEEE International Conference on Bioinformatics and Biomedicine}.\hskip
  1em plus 0.5em minus 0.4em\relax IEEE, 2017, pp. 1238--1241.

\bibitem{fournier2013meit}
P.~Fournier-Viger, E.~Mwamikazi, T.~Gueniche, and U.~Faghihi, ``{MEIT}: Memory
  efficient itemset tree for targeted association rule mining,'' in
  \emph{International Conference on Advanced Data Mining and
  Applications}.\hskip 1em plus 0.5em minus 0.4em\relax Springer, 2013, pp.
  95--106.

\bibitem{chand2012target}
C.~Chand, A.~Thakkar, and A.~Ganatra, ``Target oriented sequential pattern
  mining using recency and monetary constraints,'' \emph{International Journal
  of Computer Applications}, vol.~45, no.~10, 2012.

\bibitem{chueh2010mining}
H.~E. Chueh \emph{et~al.}, ``Mining target-oriented sequential patterns with
  time-intervals,'' \emph{International Journal of Computer Science \&
  Information Technology}, vol.~2, no.~4, pp. 113--123, 2010.

\bibitem{zhang2021tusq}
C.~Zhang, Z.~Du, Q.~Dai, W.~Gan, J.~Weng, and P.~S. Yu, ``{TUSQ}: Targeted
  high-utility sequence querying,'' \emph{arXiv preprint arXiv:2103.16615},
  2021.

\bibitem{zaki2000scalable}
M.~J. Zaki, ``Scalable algorithms for association mining,'' \emph{IEEE
  Transactions on Knowledge and Data Engineering}, vol.~12, no.~3, pp.
  372--390, 2000.

\bibitem{pasquier1999discovering}
N.~Pasquier, Y.~Bastide, R.~Taouil, and L.~Lakhal, ``Discovering frequent
  closed itemsets for association rules,'' in \emph{International Conference on
  Database Theory}.\hskip 1em plus 0.5em minus 0.4em\relax Springer, 1999, pp.
  398--416.

\bibitem{szathmary2007towards}
L.~Szathmary, A.~Napoli, and P.~Valtchev, ``Towards rare itemset mining,'' in
  \emph{The 19-th IEEE International Conference on Tools with Artificial
  Intelligence}.\hskip 1em plus 0.5em minus 0.4em\relax IEEE, 2007, pp.
  305--312.

\bibitem{aryabarzan2021neclatclosed}
N.~Aryabarzan and B.~Minaei-Bidgoli, ``{NEclatClosed}: A vertical algorithm for
  mining frequent closed itemsets,'' \emph{Expert Systems with Applications},
  vol. 174, p. 114738, 2021.

\bibitem{aryabarzan2018negfin}
N.~Aryabarzan, B.~Minaei-Bidgoli, and M.~Teshnehlab, ``{negFIN}: An efficient
  algorithm for fast mining frequent itemsets,'' \emph{Expert Systems with
  Applications}, vol. 105, pp. 129--143, 2018.

\bibitem{pei2007h}
J.~Pei, J.~Han, H.~Lu, S.~Nishio, S.~Tang, and D.~Yang, ``{H-Mine}: Fast and
  space-preserving frequent pattern mining in large databases,'' \emph{IIE
  Transactions}, vol.~39, no.~6, pp. 593--605, 2007.

\bibitem{shie2013mining}
B.~E. Shie, P.~S. Yu, and V.~S. Tseng, ``Mining interesting user behavior
  patterns in mobile commerce environments,'' \emph{Applied Intelligence},
  vol.~38, no.~3, pp. 418--435, 2013.

\bibitem{chu2008efficient}
C.~Chu, V.~S. Tseng, and T.~Liang, ``An efficient algorithm for mining temporal
  high utility itemsets from data streams,'' \emph{Journal of Systems and
  Software}, vol.~81, no.~7, pp. 1105--1117, 2008.

\bibitem{yen2007mining}
S.~J. Yen and Y.~Lee, ``Mining high utility quantitative association rules,''
  in \emph{International Conference on Data Warehousing and Knowledge
  Discovery}.\hskip 1em plus 0.5em minus 0.4em\relax Springer, 2007, pp.
  283--292.

\bibitem{tseng2010up}
V.~S. Tseng, C.~Wu, B.~E. Shie, and P.~S. Yu, ``{UP-G}rowth: an efficient
  algorithm for high utility itemset mining,'' in \emph{Proceedings of the
  16-th ACM SIGKDD International Conference on Knowledge Discovery and Data
  Mining}, 2010, pp. 253--262.

\bibitem{duong2018efficient}
Q.~H. Duong, P.~Fournier-Viger, H.~Ramampiaro, K.~N{\o}rv{\aa}g, and T.~L. Dam,
  ``Efficient high utility itemset mining using buffered utility-lists,''
  \emph{Applied Intelligence}, vol.~48, no.~7, pp. 1859--1877, 2018.

\bibitem{gan2020tophui}
W.~Gan, S.~Wan, J.~Chen, C.~M. Chen, and L.~Qiu, ``Top{HUI}: Top-$k$
  high-utility itemset mining with negative utility,'' in \emph{IEEE
  International Conference on Big Data}.\hskip 1em plus 0.5em minus 0.4em\relax
  IEEE, 2020, pp. 5350--5359.

\bibitem{tseng2015efficient}
V.~S. Tseng, C.~W. Wu, P.~Fournier-Viger, and P.~S. Yu, ``Efficient algorithms
  for mining top-$k$ high utility itemsets,'' \emph{IEEE Transactions on
  Knowledge and Data Engineering}, vol.~28, no.~1, pp. 54--67, 2015.

\bibitem{duong2016efficient}
Q.~H. Duong, B.~Liao, P.~Fournier-Viger, and T.~L. Dam, ``An efficient
  algorithm for mining the top-$k$ high utility itemsets, using novel threshold
  raising and pruning strategies,'' \emph{Knowledge-Based Systems}, vol. 104,
  pp. 106--122, 2016.

\bibitem{gan2019huopm}
W.~Gan, J.~C.~W. Lin, P.~Fournier-Viger, H.~C. Chao, and P.~S. Yu, ``{HUOPM}:
  High-utility occupancy pattern mining,'' \emph{IEEE Transactions on
  Cybernetics}, vol.~50, no.~3, pp. 1195--1208, 2020.

\bibitem{gan2018privacy}
W.~Gan, C.~W.~J. Lin, H.~C. Chao, S.~L. Wang, and P.~S. Yu, ``Privacy
  preserving utility mining: a survey,'' in \emph{IEEE International Conference
  on Big Data}.\hskip 1em plus 0.5em minus 0.4em\relax IEEE, 2018, pp.
  2617--2626.

\bibitem{abeysinghe2018query}
R.~Abeysinghe and L.~Cui, ``Query-constraint-based mining of association rules
  for exploratory analysis of clinical datasets in the national sleep research
  resource,'' \emph{BMC Medical Informatics and Decision Making}, vol.~18,
  no.~2, pp. 89--100, 2018.

\end{thebibliography}

\end{document}